\documentclass[twocolumn]{aastex631}
\usepackage{array} 
\usepackage{booktabs} 
\usepackage{makecell} 
\usepackage{graphicx} 
\usepackage{tabularx} 
\usepackage{ragged2e} 
\usepackage{multirow}

\usepackage{hyperref}
\providecommand{\footnoterule}{%
  \kern -3pt 
  \hrule width \linewidth height 0.4pt 
  \kern 6pt  
}

\begin{document}

\title{Turbulent Properties of Interplanetary Coronal Mass Ejections Observed by Solar Orbiter in the Inner Heliosphere}
\author[0000-0003-1713-119X]{Jyoti Sheoran}
\affiliation{Aryabhatta Research Institute of Observational Sciences, 
Beluwakhan, 263001, Uttarakhand, India } 
\affiliation{Department of Applied Physics, Mahatma Jyotiba Phule Rohilkhand University, Bareilly 243006, India } 
\author[0000-0002-3746-0989]{Supratik Banerjee}
\affiliation{Department of Physics, Indian Institute of Technology Kanpur, Kanpur 208016, India} 
\author[0000-0002-6954-2276]{Vaibhav Pant}
\affiliation{Aryabhatta Research Institute of Observational Sciences, 
Beluwakhan, 263001, Uttarakhand, India }

\author[0000-0003-4653-6823]{Dipankar Banerjee}
\affiliation{Indian Institute of Space Science and technology, Valiamala, Thiruvananthapuram - 695 547,Kerala, India}
\affiliation{Indian Institute of Astrophysics, Koramangala, Bangalore 560034, India}
\affiliation{Center of Excellence in Space Sciences India, IISER Kolkata, Mohanpur 741246, West Bengal, India}
\author{ M. Saleem Khan}
\affiliation{Department of Applied Physics, Mahatma Jyotiba Phule Rohilkhand University, Bareilly 243006, India }

\begin{abstract}

We investigate the turbulent properties of 12 interplanetary coronal mass ejections (ICMEs) observed by \textit{Solar Orbiter} between 0.29 and 1.0~AU. We analyze fluctuation power, spectral indices, break scales, and correlations between magnetic and velocity fluctuations ($v$–$b$) to quantify differences between ICME substructures (sheath and magnetic ejecta (ME)) and the surrounding solar wind. The ICME sheath is consistently the most turbulent region at all distances. In the solar wind, Alfv\'{e}nicity influences inertial-range scaling, resulting in either single power laws near $f^{-3/2}$ or $f^{-5/3}$, or a coexistence of both, whereas ICME substructures consistently exhibit Kolmogorov-like $f^{-5/3}$ spectra. Alfv\'{e}nicity is reduced within ICMEs, particularly in the ejecta, indicating more balanced Alfv\'{e}nic fluctuations than in the solar wind. Spectral breaks shift to higher frequencies in ICME regions, with average break frequencies of $0.53 \pm 0.35$~Hz (solar wind), $1.87 \pm 1.46$~Hz (sheath), and $1.46 \pm 1.28$~Hz (ME), reflecting differences in underlying microphysical scales. Our findings highlight distinct turbulence regimes in ICMEs compared to the solar wind and support the use of fluctuation power, spectral breaks, and $v$–$b$ correlations as effective diagnostics for identifying ICME boundaries.

\end{abstract}


\section{Introduction} \label{sec:intro}

The heliosphere is filled with the solar wind, a plasma flow that originates from the Sun and accelerates to supersonic and super-Alfv\'{e}nic speeds while continuously extending outward into interplanetary space. Comprising fluctuations of the field variables (velocity, magnetic field etc.) across a wide range of scales, the solar wind becomes a natural laboratory for the study of space plasma turbulence \citep{1995_Goldstein, 2013_Bruno, 2019_Verscharen}. Extensive studies on solar wind turbulence (see \cite{2013_Bruno} and references therein), utilizing in-situ spacecraft observations, consistently show that the magnetic power spectral densities (PSD) follow distinct power laws at different range of scales.  At the energy injection range with length scales larger than the correlation length, a $k^{-1}$ behaviour is observed (mainly in the fast solar wind), whereas a much steeper spectra with power-law exponent $< -3$ is observed at the smallest scales of dissipation \citep{1995_Tu, 2013_Bruno, 2009_Sahraoui, 2009_Alexandrova_prl}. Across the intermediate inertial scales (far from the injection and dissipation scales), power laws are associated with nonlinear cascades of energy. At scales sufficiently larger than the ion inertial length, $d_i$ (typically $\approx 100$~km near 1~AU), the magnetic power spectra follows a $k^{-5/3}$ scaling, characteristic of the energy cascade of magnetohydrodynamic (MHD) turbulence \citep{2000_McComas, 2003_Bruno, 2020_Chen, 2021_Damicis, 2024_Thepthong}. For length scales inferior to $d_i$, however, the PSD  transitions into a steeper $k^{-7/3}$ law reflecting to the current dominated dynamics of Hall MHD turbulence. 
Recently, more detailed analyses of scaling laws have emerged, revealing departures from this well-established framework. Several studies report the presence of two distinct power-law regimes within the MHD inertial range, challenging the long-held notion of a single power law \citep{2011_Wicks, 2022_Telloni, 2022_Wu, 2025_Shiladittya}.
In particular, for the fast wind close to the sun, a $k^{-3/2}$ power law is systematically observed between the $k^{-1}$ and $k^{-5/3}$ regimes. Although the main reason of such new power law remains elusive, a plausible explanation for the emergence of the $-3/2$ power law is attributed to the high Alfv\'enicity, or strong correlation between magnetic and velocity field fluctuations \citep{1971_Belcher, 2019_Kasper, 2022_Telloni, 2025_Shiladittya}.    

Alongside the continuous solar wind, the Sun occasionally releases large-scale structures called coronal mass ejections (CMEs) \citep{2012_Webb}. These CMEs typically feature a well-organized magnetic configuration, often a twisted magnetic flux rope (MFR), which facilitates their propagation through the heliosphere, where they are referred to as interplanetary CMEs (ICMEs). Compared to the ambient solar wind, ICMEs possess distinct plasma and electromagnetic properties \citep{2006_Zurbuchen}, providing a unique environment for studying the space weather phenomena. Each ICME typically consists of three distinct regions: a shock, a sheath, and the magnetic ejecta (ME). When an ICME travels faster than the surrounding solar wind, it drives a forward shock, compressing and accumulating plasma and magnetic fields in front of the ejecta, forming a sheath region. This sheath region is marked by high density, elevated proton temperature, and strongly fluctuating magnetic fields. The ME represents the magnetic driver of the ICME, and is characterized by an enhanced magnetic field, reduced magnetic fluctuations, low proton temperature, and a low plasma beta (the ratio of thermal to magnetic pressure). 
Similar to the solar wind, clear signature of turbulent fluctuations also observed in the ICMEs \citep{2013_Kilpua, 2019_Moissard}. The intricate interactions between the transient ICMEs and the ambient solar wind affect the turbulent properties of each other, thus perplexing their combined impact on heliospheric weather and the terrestrial magnetosphere. A systematic understanding of turbulence in both solar wind and ICMEs is therefore crucial. 

Despite a plethora of research on solar wind turbulence at various scales, relatively few studies are dedicated to the understanding of turbulence within ICME regions, and even fewer studies comparing them to that in the solar wind. At 1~AU,  shock-driven ICME sheaths are found to exhibit higher fluctuation power,  enhanced compressibility, spectral slopes steeper than $-5/3$, and stronger intermittency compared to the preceding solar wind \citep{2013_Kilpua, 2019_Moissard, 2020_Kilpua, 2021a_Kilpua, 2021_Sorriso, 2023_Marquez, 2025_Ruohotie}. However, at sub-ion scales, the spectral slopes are shallower than $-2.8$, suggesting a continued turbulent cascade into sub-ion scales \citep{2009_Sahraoui, 2020_Kilpua}. Within the ME region, on the other hand, the transverse fluctuations are often dominant and less-steep power-laws, with a spectral index of approximately $-1.5$ in the MHD range and $-2$ in the sub-ion range, are observed \citep{1998_Leamon, 2008_Hamilton}. However, a recent study at $1$ AU reports a consistent ${-5/3}$ scaling for the ME region along with zero influence of the background magnetic field \citep{2024_Shaikh}. Furthermore, a comparative study, between ICMEs at 0.47 and 1.08~AU, observes an increase of the spectral slope with heliocentric distance along with a decreasing compressibility in the sheath region \citep{2020a_Good}. 
Interestingly, unlike the non-ICME solar wind, the MHD range power spectra of the ME part
follow a $-5/3$ power law at all radial distances (less than and at 1 AU), but steepen to values consistently below $-3$ towards the kinetic scales. The spectral break between the inertial and sub-ion spectra for ICME ejecta are found to lie closer to the ion inertial length than to the ion cyclotron frequency \citep{2023_Good, 2024_Shaikh}. Furthermore, the nature of fluctuations (Alfv\'{e}nic or non-Alfv\'{e}nic) significantly influences turbulence properties. In both the solar wind and ICMEs, Alfvénicity is commonly quantified using the cross helicity ($\sigma_c$), which measures the dominance of unidirectional Alfvén waves through correlations or anti-correlations between magnetic and velocity fluctuations \citep{2019_Stabsby}. High $\sigma_c$ values (near $\pm1$) indicate the presence of unidirectional Alfv\'{e}n waves, associated with an imbalance in power between waves propagating parallel or antiparallel to the mean magnetic field. Conversely, low $\sigma_c$ values (near zero) suggest balanced, non-Alfv\'{e}nic turbulence. Previous studies have shown that ICME sheaths and the ME exhibit reduced $\sigma_c$ compared to the ambient solar wind \citep{2022_Good, 2020a_Good, 2015_Wien}. Sheaths—formed by compressed solar wind ahead of ICMEs—typically retain moderate $\sigma_c$ values, reflecting a combination of the anti-sunward dominance of the upstream solar wind and locally generated balanced turbulence. Flux ropes, on the other hand, tend to exhibit the lowest $\sigma_c$ values due to their closed magnetic topology, which allows bidirectional propagation of Alfv\'{e}nic fluctuations and results in a mixing of sunward and anti-sunward wave populations originating from the corona.

Despite several efforts to characterize ICME turbulence, many aspects remain poorly understood. Key questions include: How do the spectral properties of ICME regions (sheath and ME) evolve with radial distance in the inner heliosphere? How does the nature and strength of turbulence within ICME regions (sheath and ME) compare to that of the ambient solar wind? How does the turbulence of a non-ICME solar wind is modified by the visit of an ICME? To address these questions, using data of the Solar Orbiter (SolO) spacecraft, we perform a systematic comparison of the PSD of magnetic field fluctuations between the sheath, the ME region and the ambient solar wind before and after the ICME. Our study focuses on key turbulence characteristics, including fluctuation power, spectral indices, break scales, and correlation coefficients between magnetic and velocity field fluctuations, to quantify differences between ICME substructures and the surrounding solar wind. Establishing such distinctions is crucial because the identification of ICME in in-situ data remains partly subjective, as no single physical parameter provides a definitive detection \citep{2006_Zurbuchen}. In a broader view, this study aims to explore whether ICMEs exhibit a distinct nature of turbulence and whether this distinction can aid in their reliable identification within the solar wind.

The paper is organized as follows: Section \ref{sec:data} describes the observational data. Section \ref{sec:analysis} outlines the analysis methods. Section \ref{sec:results} presents the results and key findings. Finally, Section \ref{sec:summary} summarizes the study and concludes.

\begin{figure*}[h!]
    \centering
    \includegraphics[width=\textwidth]{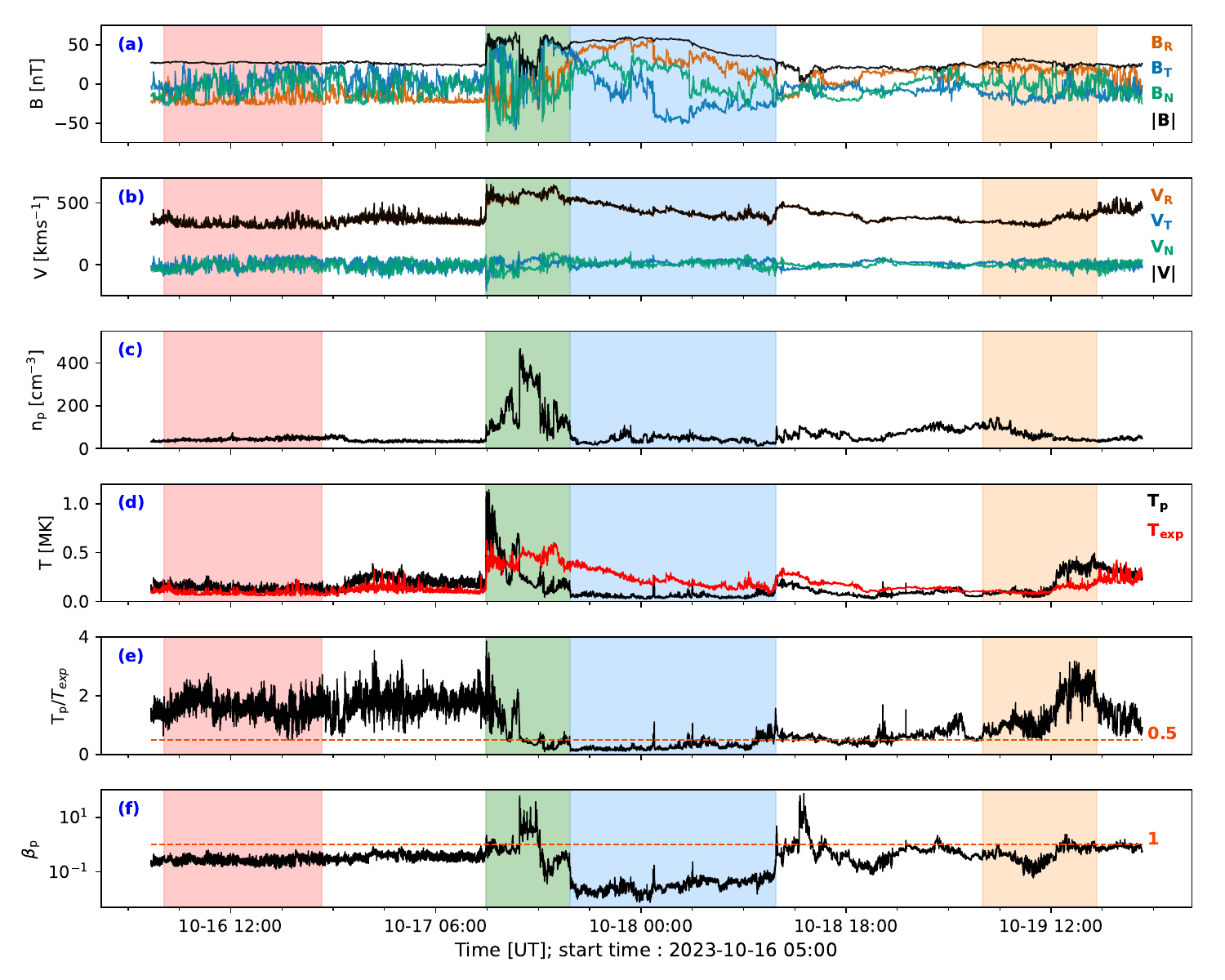}
\caption{The magnetic field and plasma data for an ICME observed by 
SolO spacecraft on October 17, 2023. Panel (a): Magnetic field components and magnitude, Panel (b): Proton velocity components and magnitude,  Panel (c): Proton density,   Panel (d): Proton temperature ($T_p$, blue points) and expected proton temperature ($T_{\text{exp}}$) derived from the $V_{\text{sw}}$ relation (red points),  Panel (e): Ratio of $T_p/T_{\text{exp}}$ ( dashed red line corresponds to $T_p/T_{\text{exp}} = 0.5$),  Panel (f): Proton plasma beta ($\beta_p$)( dashed red line corresponds to $\beta_p = 1$).  Colored areas represent four selected regions: light salmon pink – SW1, light green – sheath, light blue – ME, and light peach – SW2.
}
    \label{fig:icme_231017}
\end{figure*}

\section{Observational Data} \label{sec:data}

We used in-situ plasma and magnetic field measurements from the SolO spacecraft for this study. Magnetic field data were obtained from the SolO magnetometer (MAG; \citealt{2020_Horbury}), which operates in two modes: Normal mode, recording data at 8 to 16 vectors per second, and Burst mode, providing higher resolution at 64 to 128 vectors per second. The plasma data were sourced  from the Solar Wind Analyser \citep[SWA;][]{2020_Owen} suite, with a typical time resolution of 4 seconds provided by the SWA-Proton Alpha Sensor (PAS).

The Helio4Cast ICMECAT database\footnote{\href{https://helioforecast.space/icmecat}{https://helioforecast.space/icmecat}} \citep{2017_Mostl} offers a regularly updated catalog of ICMEs observed by SolO. From this database, we selected ICME events spanning April~2021 to March~2024. Note that April~2021 marks the earliest period for which continuous, high-cadence plasma measurements from the PAS instrument are consistently available, and earlier events were therefore excluded. The primary selection criteria required continuous magnetic-field and plasma measurements throughout the ICME interval, with a negligibly small number of data gaps  ($\lesssim 1\%$) , as well as sufficiently long pre- and post-ICME solar-wind intervals to characterize the background turbulence. Events involving consecutive or overlapping ICMEs were excluded to avoid ambiguity in boundary identification and spectral analysis. 
Based on these criteria, we identified 12 ICMEs suitable for the present study. These events span heliocentric distances from 0.29 to 1.01~AU, providing broad radial coverage within the inner heliosphere, and include both fast and slow ICMEs. For each event, ICME boundaries were initially taken from the Helio4Cast ICMECAT catalog and then refined through visual inspection, as follows. For each ICME event, we plotted the magnetic field data using the SolO/MAG Normal Mode one-minute resolution data and the plasma parameters using SWA-PAS data. Figure~\ref{fig:icme_231017} depicts an ICME that started on October 17, 2023, at 10:24 UT, as observed by SolO at a heliocentric distance of 0.37 AU. The ICME regions, including the sheath and ME, were identified based on the characteristics observed in both the magnetic field and plasma data. The sheath region (highlighted in light green in Figure~\ref{fig:icme_231017}) is characterized by a strongly fluctuating (turbulent) magnetic field, high density, elevated temperature, and plasma beta larger than one. The ME (highlighted in light sky blue color in Figure~\ref{fig:icme_231017}) is identified by a notable enhancement and/or rotation of the magnetic field, a low plasma beta ($\beta_p < 1$), a declining velocity profile indicative of expansion, and a depressed proton temperature ($T_p/T_{\text{exp}} < 0.5$; $T_{\text{exp}}$: derived from the empirical $V_{\text{sw}} - T_p$ correlation \citep{1986_Lopez, 2005_Liu}. 
Additionally, background solar wind regions preceding (SWA) and following (SWB) the ICME, shown in salmon pink and peach, respectively, were selected for comparative analysis, ensuring that these regions were sufficiently long to maintain statistical accuracy.  SWA regions were selected to be reasonably stationary, while SWB regions were carefully examined to exclude any rotations in the magnetic field components, thereby avoiding contamination from ICME-affected solar wind. This approach accounts for the variability in post-ICME solar wind conditions and the inherent subjectivity in defining ICME boundaries. In the appendix, we present the in situ measurements and identified regions for all ICME events considered in this study (Figure~\ref{fig:rawdata_a}) alongside a table listing the event dates and distances (Table~\ref{tab:icme_events}).

\section{Analysis and Methods} \label{sec:analysis}

We characterized turbulence by computing (i) the magnetic power spectral density (PSD) and (ii) the Pearson correlation coefficients between magnetic and velocity field fluctuations. We computed the PSD using SolO/MAG Burst mode data when available, or Normal mode otherwise, with time resolutions from $\sim$ 0.0156 to $\sim$0.127 seconds for the 12 selected events. For the correlation analysis, we downsampled the magnetic data to match the plasma measurement cadence of 4 seconds. All data sets were uniformly sampled by interpolating data gaps prior to analysis to ensure consistency in the time series.

Since \( V_A \ll V \) in our all selected solar wind and ICME regions, where \( V_A \) and \( V \) are the Alfvén speed and flow speed, respectively, we can apply Taylor’s hypothesis \citep{1938_Taylor}, which allows spatial scales \( \ell \) to be inferred from temporal measurements \( \tau \) using the relation \( \ell = V \tau \) \citep{2014_Howes}. Therefore, we define the fluctuations in the \( i \)-th component (where \( i = r, t, n \) in the RTN coordinate system) of the magnetic and velocity fields as:
\(
\delta B_i(t, \tau) = B_i(t + \tau) - B_i(t), \quad and \quad   \delta V_i(t, \tau) = V_i(t + \tau) - V_i(t)
\).

The magnetic power spectral density (PSD) is defined as 
\(\mathrm{PSD} = \hat{B}_i \hat{B}_i^*\), 
with summation over \(i = r, t, n\), where \(\hat{B}_i\) represents the Fourier transform of the magnetic field component \(B_i\), computed using the fast Fourier transform (FFT). The PSD typically exhibits a power-law dependence, \(\sim f^\alpha\), which characterizes the turbulent cascade. To quantify this, we determine the spectral indices (\(\alpha\)) by performing linear fits to the magnetic power spectra in logarithmic space, separately within the inertial and kinetic ranges for each region. Although the inertial range is often modeled using a single power-law fit, recent studies of solar wind turbulence \citep{2025_Shiladittya} suggest the presence of an additional break within this range. To account for this, we apply the Kolmogorov--Smirnov (K--S) test to identify a potential break point (\(f_{b_i}\)) within the inertial range. Specifically, we calculate the K--S distances (\(\mathrm{K\text{-}S}_1, \mathrm{K\text{-}S}_2\)) between the original and fitted spectra for two subranges split at \(f_{b_i}\), and select the break point that minimizes the maximum of the two distances, i.e., \(\max(\mathrm{K\text{-}S}_1, \mathrm{K\text{-}S}_2)\). We then compare this value to the K--S distance (\(\mathrm{K\text{-}S}_3\)) obtained from fitting a single power law to the full inertial range. If \(\max(\mathrm{K\text{-}S}_1, \mathrm{K\text{-}S}_2)_{f_{b_i}} < \mathrm{K\text{-}S}_3\), the inertial range is considered better described by a double power law with a break at \(f_{b_i}\); otherwise, a single power law provides a better fit. It is worth noting that, to ensure a consistent comparison of turbulence properties across ICME regions and the ambient solar wind at inertial scales, we restrict our analysis to the frequency range \(10^{-2} \leq f \leq 5 \times 10^{-1}\,\mathrm{Hz}\). The lower limit is determined by ICME durations, while the upper limit is chosen to remain within the inertial range, beyond which the spectra steepen toward the dissipation range.

Furthermore, we identify the high-frequency spectral break (\(f_b\)), which marks the transition from the MHD inertial range to the ion kinetic regime, and compare it with the characteristic frequencies associated with the ion inertial length (\(d_i\)) and the ion gyroradius (\(\rho_i\)). The frequencies corresponding to these scales are calculated using the following expressions:

\begin{equation}
f_{d_i} = {e \langle V \rangle} \sqrt{\frac{\mu_0 \langle n \rangle}{m_i}}
\end{equation}

\begin{equation}
f_{\rho_i} = \frac{e \langle V \rangle \langle B \rangle}{ \sqrt{2 k_B \langle T_i \rangle m_i}}
\end{equation}
Here, \( T_i \), \( m_i \), \( e \), and \( V \) denote the proton temperature, mass, charge, and speed, respectively. The angle brackets indicate time-averaged values over the interval.

To further examine the nature of fluctuations, we compute the Pearson correlation coefficients between magnetic and velocity field fluctuations as:

\begin{equation}
\rho_{\delta B_i, \delta V_i} 
= \frac{\mathrm{Cov}(\delta B_i, \delta V_i)}{\sigma_{\delta B_i} \, \sigma_{\delta V_i}} ,
\end{equation}
where $\mathrm{Cov}(\delta B_i, \delta V_i) = \langle \delta B_i \, \delta V_i \rangle$ is the covariance between $\delta B_i$ and $\delta V_i$. These coefficients quantify the degree of correlation between the two fields and provide insights into the Alfv\'{e}nic nature of the turbulence.

 \begin{figure*}
    \centering
    \includegraphics[width=\textwidth]{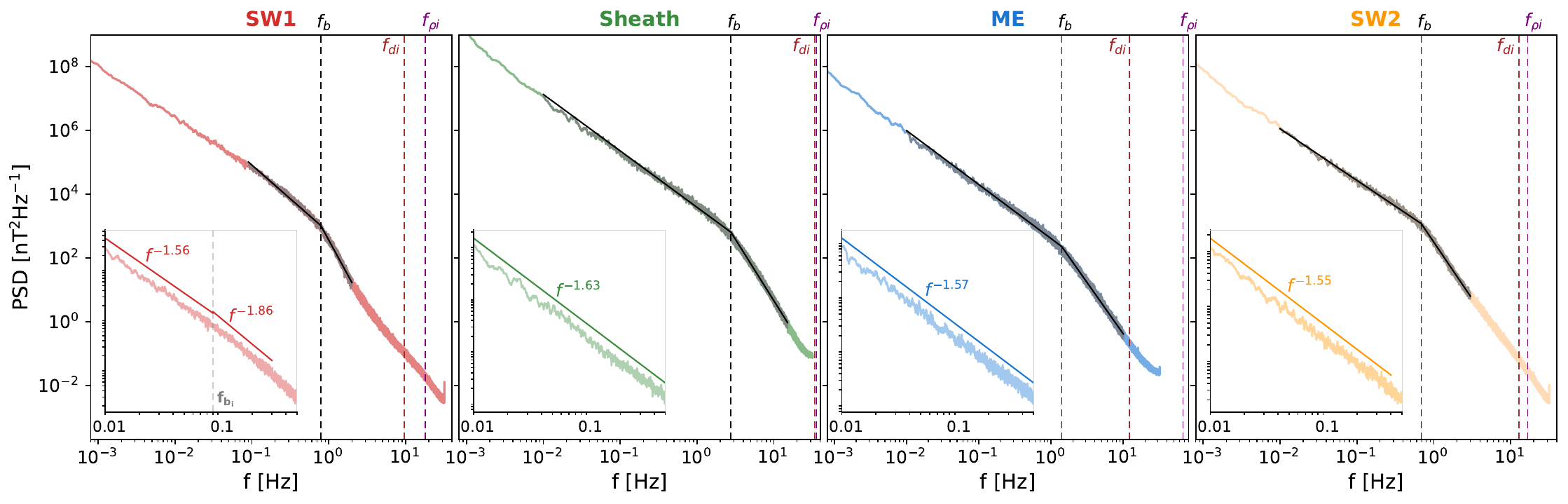}
    \caption{Magnetic PSD traces for the selected regions of the ICME observed on 2023 October 17, smoothed using a running mean window. The insets show the fitted PSDs within the inertial range ($10^{-2}$– $ 5 \times 10^{-1}$ Hz) for each region.}
    \label{fig:psd_20231017}
\end{figure*}

\section{Results and Discussion}\label{sec:results}


\subsection{Turbulence properties of an ICME at 0.37 AU: A Typical Case}\label{sec:psd}
Figure~\ref{fig:icme_231017} shows an ICME observed by SolO on 2023 October 17 at a heliocentric distance of 0.37~AU. Figure~\ref{fig:psd_20231017} presents the smoothed trace PSDs as a function of spacecraft-frame frequency for four selected regions (SW1, SH, ME, and SW2). We apply a running-average window to the PSDs to enhance the visibility of spectral features. The results indicate that the sheath region exhibits significantly higher fluctuation power compared to the SW1, ME, and SW2 regions, identifying it as the most turbulent. We further fit the PSDs within the inertial range ($10^{-2} - 5 \times 10^{-1}\,\mathrm{Hz}$) for these regions, as shown in the insets of Figure~\ref{fig:psd_20231017}. In the SW1 region, we identify a distinct spectral break (\(\mathrm{f}_{b_i}\)) within the inertial range, indicated by a vertical dashed line. Below this break, the spectrum follows a slope close to $-3/2$, transitioning to a $-5/3$ scaling at higher frequencies. In contrast, the sheath, ME, and SW2 regions each follow a single power-law scaling, with spectral indices of $-1.63$, $-1.57$, and $-1.58$, respectively. In the context of isotropic turbulence, an $f^{-5/3}$ power-law is typically associated with Kolmogorov-like turbulence, where energy cascades through a hierarchy of eddy fragmentation under strong nonlinear interactions \citep{Kolmogorov_1941}. Conversely, an $f^{-3/2}$ spectral slope is often linked to weaker turbulence, dominated by sporadic collisions of Alfv\'{e}nic wave packets \citep{1964S_Iroshnikov, 1965_Kraichnan}. However, in space plasmas, the presence of magnetic fields introduces anisotropy, permitting both $-5/3$ and $-3/2$ scalings under different conditions  \citep{1995_Goldreich, 1997_Goldreich, 2015_Chandran}. The observations of this  event suggest that in the SW1 region, turbulence is less developed at larger scales and becomes more fully developed at smaller scales. The arrival of the ICME appears to enhance turbulent mixing, particularly within the sheath region, leading to a more developed turbulent state characterized by a spectral slope close to $f^{-5/3}$. In contrast, the ME and SW2 regions exhibit shallower spectral slopes than the sheath, indicative of relatively weaker turbulence dominated by Alfvénic fluctuations.

Further, ion-scale spectral breaks are evident in the spectra at higher frequencies, marking the transition from the MHD inertial range to the sub-ion range, where energy is dissipated through kinetic processes. The gray overlay in Figure~\ref{fig:psd_20231017} highlights the frequency range spanning from the start of the inertial range (taken as $10^{-2}$~Hz in this study) to the end of the dissipation range. If a spectral break is identified within the inertial range for a given region, the gray spectrum begins at the break frequency $f_{bi}$ (as seen, for example, in the SW1 region). We apply a piecewise linear fit to the overlaid gray spectra using a chi-square minimization method, treating the breakpoint as a free parameter. The resulting fit is shown by the solid black line, and the corresponding breakpoint frequency is indicated by a vertical dashed black line. For reference, the vertical dashed brown and purple lines indicate the frequencies associated with the ion inertial length ($f_{d_i}$) and ion gyroradius ($f_{\rho_i}$), respectively. We find that the spectral breaks in the sheath and ME regions occur at higher frequencies compared to those in the surrounding solar wind regions (SW1 and SW2). Moreover, in all regions (SW1, sheath, ME, and SW2), the spectral breaks occur closer to the ion inertial length. This tendency is  particularly pronounced in the low-$\beta$ ME regions, consistent with \citet{2014_Chen}, who showed that in low-$\beta$ solar wind plasma, spectral breaks systematically coincide with the ion inertial length.

\begin{figure*}
    \centering
    \includegraphics[width=\textwidth]{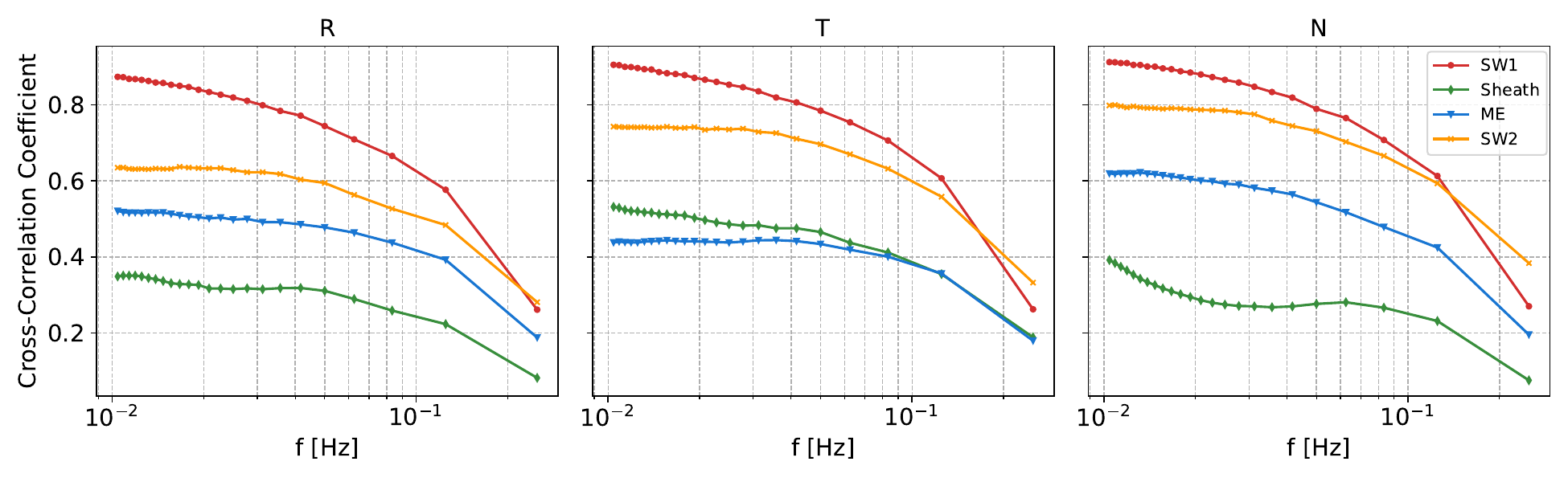}
\caption{Absolute cross-correlation coefficients between magnetic and velocity field fluctuations in the R, T, and N components, plotted as a function of frequency ($1/\tau$), for the SW1, sheath, ME, and SW2 regions of the ICME event on 2023 October 17. Here, $\tau$ is the fluctuation timescale.}
    \label{fig:cc_20231017}
\end{figure*}

Figure~\ref{fig:cc_20231017} shows the absolute values of  Pearson correlation coefficients between the magnetic and velocity field fluctuations for the R (\texttt{cc\_r}), T (\texttt{cc\_t}), and N (\texttt{cc\_n}) components for this ICME event. In the frequency range $10^{-2} < f < 10^{-1}$~Hz, the solar wind regions (SW1 and SW2) exhibit strong correlations across all components, indicative of dominant Alfv\'{e}nic fluctuations at these scales. In contrast, both the sheath and ME regions display reduced correlations relative to the solar wind, reflecting a weaker Alfv\'{e}nic nature in these regions.
Studies by \citet{2020b_Good, 2022_Good} have quantified the Alfv\'{e}nicity within ICMEs using the normalized cross helicity ($\sigma_c$), which measures the dominance of unidirectional Alfv\'{e}n waves. 
Their results indicate that both the sheath and ME regions generally exhibit lower values of $\sigma_c$ compared to the ambient solar wind, where $\sigma_c$ is typically higher. 
The reduced correlations observed in the sheath and ME in our study may therefore reflect a more balanced distribution of wave power propagating parallel and antiparallel to the mean magnetic field compared to the surrounding solar wind. Further, it can be seen that at higher frequencies, the correlation coefficients decrease in all regions, suggesting a transition to more fully developed turbulence at smaller scales, where fluctuations become increasingly balanced and less Alfv\'{e}nic.

\subsection{ Statistical Characteristics of Turbulence with Radial Distance}

The in-situ measurements and identified regions for all ICME events analyzed in this study are shown in Figure~\ref{fig:rawdata_a} in the Appendix. Figure~\ref{fig:psds_all} presents the trace PSDs for the selected regions of all events, with the spectra color-coded according to their radial distances. It is evident that, at all heliocentric distances, the sheath region exhibits enhanced fluctuation power compared to both the ambient solar wind and ME regions. Previous studies \citep{2025_Ghuge, 2019_Moissard, 2021a_Kilpua, 2017_Kilpua} have reported a higher level of magnetic fluctuations in ICME sheaths at 1~AU. Our results extend these findings to distances below 1~AU, demonstrating that the sheath is the most turbulent region relative to the surrounding solar wind and ME regions. The higher fluctuation power observed in the sheath regions is likely a consequence of their nature as compressed zones between the shock front and the magnetic ejecta. These regions experience strong shock compression and solar wind pile-up, which can amplify magnetic field fluctuations. Moreover, Figure~\ref{fig:psds_all} shows a clear decline in fluctuation power with increasing distance in all regions, consistent with previous reports for the solar wind \citep{2020_Chen} and ICME ejecta \citep{2023_Good}. This radial trend reflects the weakening of the background magnetic field and the decreasing amplitude of turbulent fluctuations across all regions (solar wind, sheath, and ME) with increasing heliocentric distance.

\begin{figure*}
    \centering
    \includegraphics[width=0.8\textwidth]{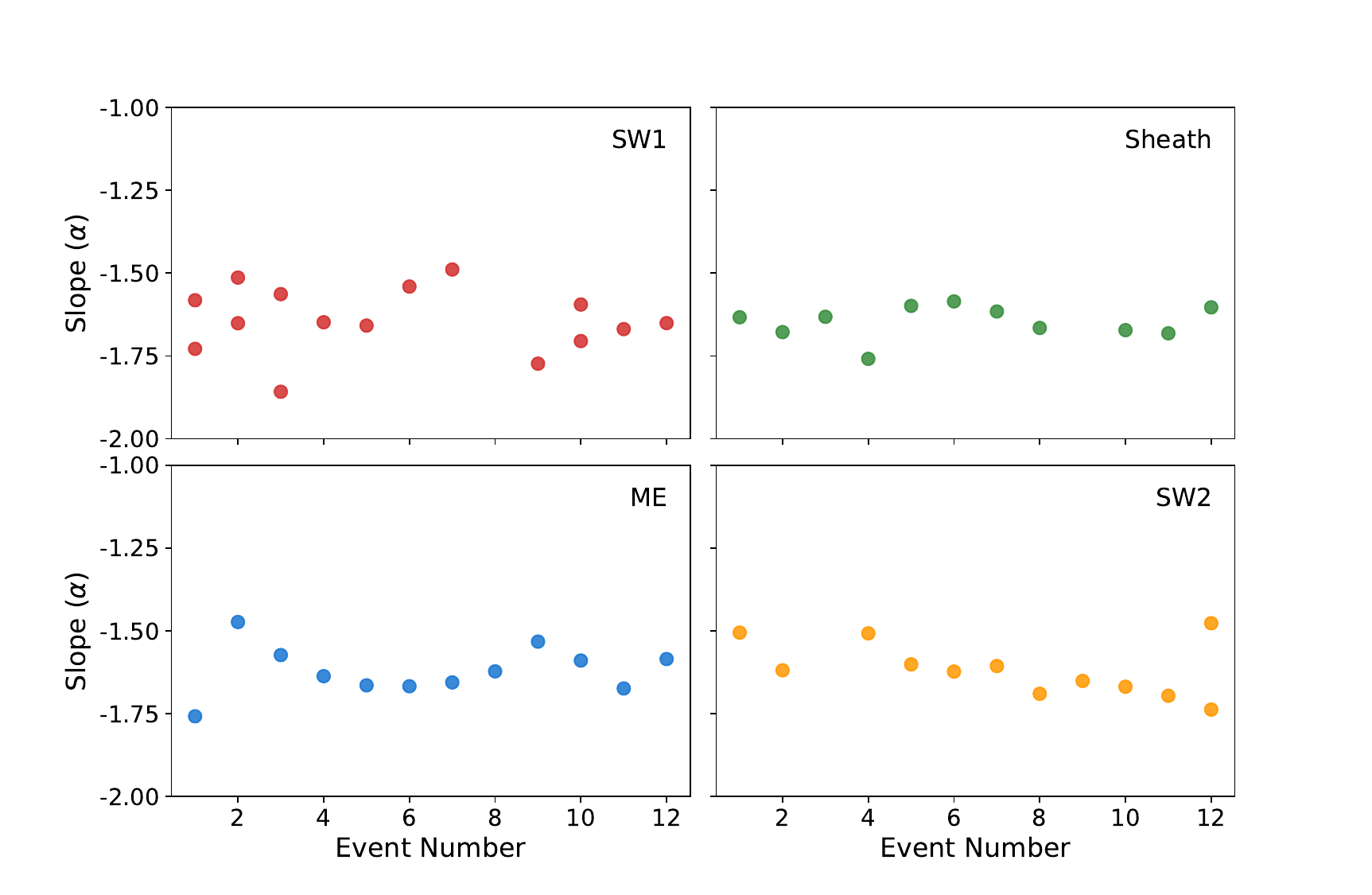}
\caption{Inertial range spectral slopes for the SW1, sheath, ME, and SW2 regions plotted against event numbers listed in Table~\ref{tab:icme_events}. While the SW1 and SW2 regions occasionally exhibit double power-law behavior with coexisting $-3/2$ and $-5/3$ slopes, the sheath and ME regions consistently display a single spectral slope close to $-5/3$ across the inertial range.}
    \label{fig:spectral_slope}
\end{figure*}

Figure~\ref{fig:spectral_slope} shows the inertial-range spectral slopes for the SW1, sheath, ME, and SW2 regions against the event numbers listed in Table~\ref{tab:icme_events}. It is worth noting that the event numbers are arranged in order of increasing heliocentric distance. The analysis reveals distinct turbulence regimes within the solar wind surrounding ICMEs (SW1 and SW2). Notably, the solar wind turbulence spectra occasionally exhibit two sub-regimes with spectral indices close to $-3/2$ and $-5/3$ coexisting within the inertial range, consistent with recent findings by \cite{2025_Shiladittya}.  In other solar wind regions, a single power-law spectrum dominates, displaying either $f^{-3/2}$ or $f^{-5/3}$ scaling. These variations will be discussed in more detail later in this section. Across all events, the ICME sheath consistently exhibits a single inertial-range power-law with a slope close to $-5/3$, indicating a well-developed turbulent cascade. This finding aligns with earlier reports of enhanced and fully developed turbulence within sheaths at 1~AU \citep{2021a_Kilpua, 2021_Sorriso}. The ME regions also predominantly follow an $f^{-5/3}$-like scaling, characteristic of strong, Kolmogorov-like turbulence. However, one exceptional case (Event~9) exhibits a shallower, $f^{-3/2}$-like spectrum in the ME, indicating the presence of weaker, Alfv'{e}nic-driven turbulence. This event is discussed separately in a later subsection. The present results are  in line with recent statistical studies \citep{2024_Shaikh, 2023_Good}, which reported inertial-range slopes between $-3/2$ and $-5/3$ in ICME ejecta, with averages closer to the Kolmogorov value. Furthermore, no systematic trend in the spectral indices of the sheath or ME regions is observed with heliocentric distance. This suggests that turbulence within ICME substructures remains well developed throughout the inner heliosphere, from as close as 0.29~AU to 1~AU.

\begin{figure*}
    \centering
    \includegraphics[width=0.8\textwidth]{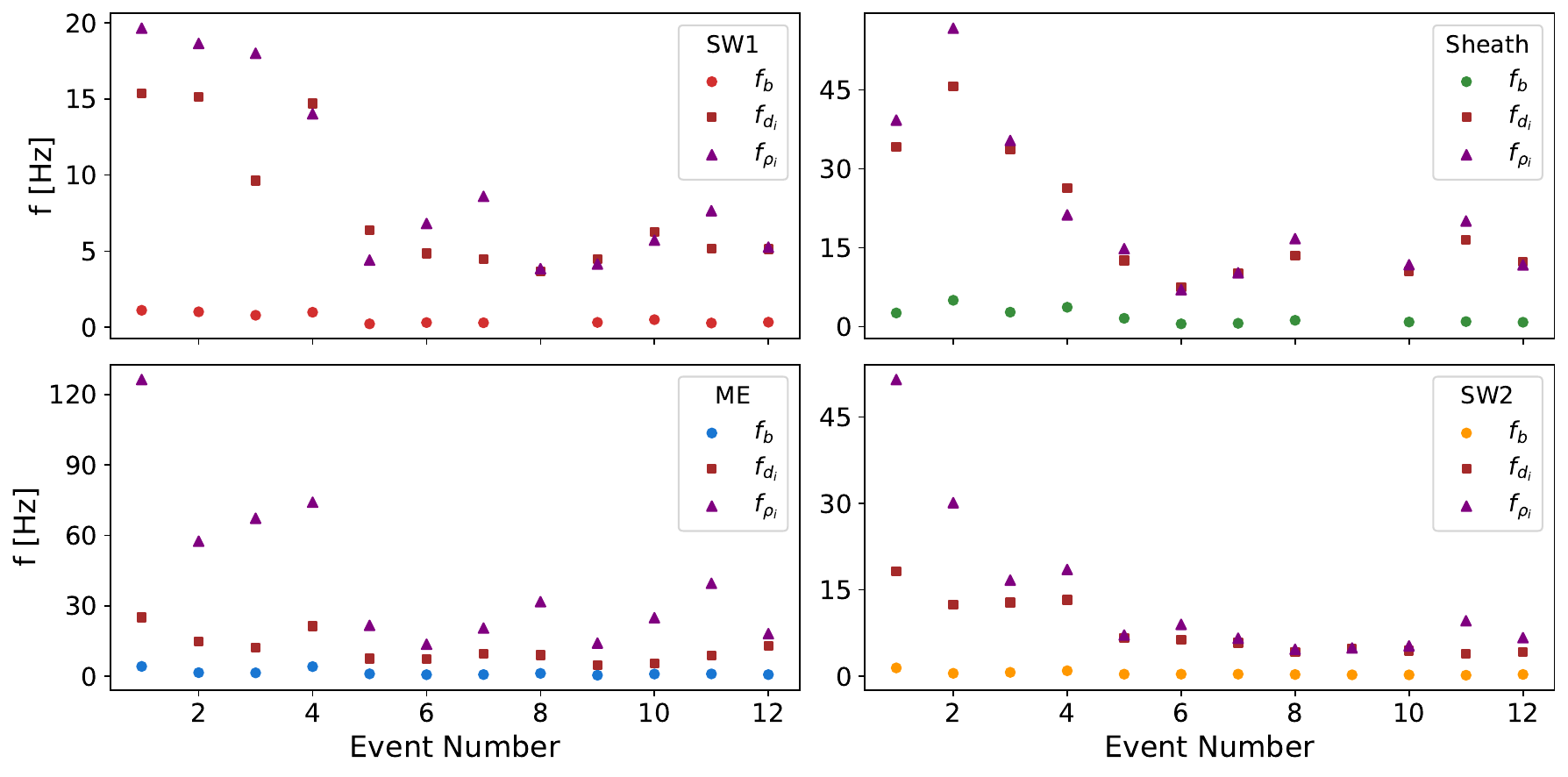}
\caption{$f_{b}$, $f_{d_i}$, and $f_{\rho_i}$ plotted against event numbers from Table~\ref{tab:icme_events}, which are ordered by increasing heliocentric distance, for the SW1, sheath, ME, and SW2 regions across all analyzed events.}
    \label{fig:bp_radial}
\end{figure*}

Our analysis reveals that the spectral breaks in the sheath and ME regions occur at higher frequencies than those in the surrounding solar wind (SW1 and SW2), a trend that remains consistent across all events. For the analyzed intervals, the average break frequencies are $0.53 \pm 0.35$~Hz for the solar wind (SW1 and SW2), $1.87 \pm 1.46$~Hz for the sheath, and $1.46 \pm 1.28$~Hz for the ME. This systematic shift toward higher break frequencies within ICMEs likely reflects differences in plasma microphysical scales relative to the ambient solar wind. Figure~\ref{fig:bp_radial} shows the radial evolution of $f_{b}$, $f_{d_i}$, and $f_{\rho_i}$ for each region across all events. A clear decrease in $f_{b}$ with increasing heliocentric distance is observed in all regions (SW1, sheath, ME, and SW2), reflecting the radial decline of the ion-scale characteristic frequencies ($f_{d_i}$ and $f_{\rho_i}$). Similar trends have been reported previously for the solar wind \citep{2014_Bruno, 2020_Duan} and ICME ejecta \citep{2023_Good}. Moreover, across all regions, the spectral breaks lie near the ion inertial length, a feature more evident in low-$\beta$ ME regions. This behavior aligns with earlier findings that, in low-$\beta$ plasma, spectral breaks occur closer to the ion inertial scale than to the ion gyroscale \citep{2014_Chen, 2023_Good}.  In the dissipation range, the spectral indices for both the solar wind and ICME regions (sheath and ME) typically lie between $-2.5$ and $-4$, in agreement with earlier findings. A detailed investigation of the dissipation-range properties, however, lies beyond the scope of the present study.

Figure~\ref{fig:cc_all} presents the Pearson correlation coefficients between the fluctuations in the radial components of the magnetic and velocity fields (\texttt{cc\_r}) for all events. Only the radial component is shown, as the tangential and normal components exhibit similar trends. Overall, across all heliocentric distances, the ICME sheath and ME regions consistently show reduced correlations compared to the surrounding solar wind (SW1 and SW2), with one exception (Event~9), where the ME region exhibits a higher correlation than the adjacent solar wind intervals. While the sheath regions generally exhibit weaker correlations than the solar wind, a few cases retain moderately high values. This behavior likely reflects their origin as compressed solar wind plasma accumulated ahead of the ICME during its interplanetary propagation \citep{2022_Good}. These regions may preserve remnants of the solar wind’s Alfv\'{e}nic character, although interactions such as velocity shear \citep{2023_Soljento} or shock-driven turbulence \citep{2023_Sishtla} can promote mixing and reduce overall Alfv\'{e}nicity. The ME regions, in contrast, predominantly exhibit very low correlations across most events, with the single exception of Event~9, where strong correlations are observed within the ejecta. The generally weak correlations in ME regions can be attributed to their closed magnetic field topology, which enables the bidirectional propagation of Alfv\'{e}nic fluctuations \citep{2020b_Good}. The resulting balance between sunward and anti-sunward wave populations reduces directional asymmetry and thus weakens the observed correlations. \citet{2016_Li} also reported a lower occurrence rate of Alfv\'{e}n waves within ICMEs relative to the ambient solar wind within 4.75~au, supporting the reduced Alfv\'{e}nic nature of these regions. However, the relatively high correlations observed in the one exceptional ME case need to be investigated in detail to understand the underlying conditions that may support enhanced Alfv\'{e}nicity within the ejecta.

To explain, the variable inertial-range scaling of solar wind region, the solar wind has recently been categorized into three types based on solar wind speed and  Alfv\'{e}nicity (\cite{2024_Ervin} and references therein): slow solar wind (SSW), fast solar wind (FSW), and Alfv\'{e}nic slow solar wind (ASSW). To aid interpretation, we have similarly classified our solar wind regions (SW1 and SW2) into SSW, FSW, and ASSW, following the criteria outlined in Table~\ref{tab:sw_class}.   In SSW regions, we observe a single power law with steeper spectra at larger heliocentric distances.  In contrast, the ASSW and FSW regions exhibit either a single power-law scaling close to $f^{-3/2}$ or $f^{-5/3}$, or a double power-law behavior, with $f^{-3/2}$ at lower frequencies and $f^{-5/3}$ at higher frequencies. The spectral indices in both ASSW and FSW regions show no clear trend with heliocentric distance. The exact values of the spectral indices for SSW, FSW, and ASSW regions are listed in Table~\ref{tab:sw_class}. Previous studies report that the solar wind near the Sun typically follows a $f^{-3/2}$ scaling, which steepens to $f^{-5/3}$ farther out in the heliosphere \citep{2020_Chen, 2021_Shi, 2023_Sioulas}. Additionally, regions with low Alfv\'{e}nic content have been found to significantly steepen with distance, while highly Alfv\'{e}nic regions retain their near-Sun scaling \citep{2023_Sioulas}. This could possibly explain the steepening observed in our SSW regions, whereas ASSW and FSW show no clear trend. Furthermore, a recent study by \citet{2025_Shiladittya} reported two distinct sub-regimes in the FSW magnetic power spectrum during low solar activity, which are not consistently observed during high solar activity. Since our regions correspond to the rising phase of solar activity, this may account for the observed variations in spectral indices in FSW and ASSW.

\subsubsection{Expectional Case: Event 9}\label{sec:event9}

Unlike most events where the ME region exhibits an $f^{-5/3}$ scaling and only a weak $\mathbf{v}$--$\mathbf{b}$ correlation within the inertial range, an exceptional case (event~9) displays an $f^{-3/2}$-like power-law behavior accompanied by enhanced $\mathbf{v}$--$\mathbf{b}$ correlation relative to the surrounding solar wind. \citet{2024_Shaikh} also reported a similar $f^{-3/2}$ spectral scaling for an ICME flux rope at 1~AU, attributing it to Iroshnikov--Kraichnan (IK) turbulence \citep{1964_Iroshnikov, 1965_Kraichnan}. However, as emphasized in \citet{2025_Shiladittya}, the classical IK phenomenology is strictly applicable to balanced MHD turbulence and cannot account for the $-3/2$ spectrum under strong $\mathbf{v}$--$\mathbf{b}$ correlation. Therefore, the observed $f^{-3/2}$ regime in this ICME ejecta is more plausibly associated with anisotropic turbulence resulting from strong $\mathbf{v}$--$\mathbf{b}$ alignment \citep{1995_Goldreich, 2006_Boldyrev}. Furthermore, the strong $\mathbf{v}$--$\mathbf{b}$ correlation observed in this ICME ejecta could be explained by the scenario proposed by \citet[see their Figure~6]{2020b_Good}, where ICMEs show higher $|\sigma_c|$ values at larger crossing distances from the flux rope axis and within the rope legs. This enhancement is likely due to local interactions with the solar wind and reconnection-driven opening of magnetic field lines, causing the ICME plasma to acquire solar-wind-like properties. 

To further investigate this event, we identified its near-Sun white-light counterpart CME. The CME first appeared in STEREO/COR2 on 2021~May~2 at 12:23~UT and in SOHO/LASCO~C2 at 12:24~UT at the east limb, propagating mainly eastward with a projected linear speed of 285~km~s$^{-1}$ (LASCO CME Catalog). We performed Graduated Cylindrical Shell \citep[GCS;][]{2006_Thernisien} modeling using STEREO-A/COR2 and SOHO/LASCO~C2 and C3 images, obtaining a propagation direction of $-133^{\circ}$ longitude and $4.5^{\circ}$ latitude (HEEQ) and a half-width  of $36.6^{\circ}$. A linear fit to the GCS apex heights between 4.78 and 7.86~$R_{\odot}$ (from 12:38 to 14:23~UT) yielded a radial speed of 350~km~s$^{-1}$, indicating that this is a narrow and slow CME. 

At the time of the ICME observation, \textit{SolO} was located at 0.91~au, $-97.4^{\circ}$ longitude, and $-0.3^{\circ}$ latitude (HEEQ), i.e., 35.6$^{\circ}$ west of the CME’s central propagation direction. Given the CME’s $36.6^{\circ}$ half-width, this geometry suggests that the flank  of the CME crossed the \textit{SolO} during its encounter. This event, therefore, observationally supports the scenario proposed by \citet{2020b_Good}, in which ICME legs exhibit a stronger $\mathbf{v}$--$\mathbf{b}$ correlation and, consequently, higher $|\sigma_c|$.

\section{Summary and Conclusion}\label{sec:summary}

We analyzed the turbulent properties of 12 ICMEs observed by the \textit{SolO} between 0.29 and 1.0~au, focusing on magnetic field power spectra, spectral indices, break scales, and cross-field correlations between magnetic and velocity fluctuations across the sheath, ME, and surrounding solar wind regions. Our results show that the ICME sheath consistently exhibits the highest fluctuation power compared to both the solar wind and the ME, indicating that the sheath is the most turbulent region throughout the observed heliocentric range. The enhanced fluctuation power in the sheath likely arises from strong compression between the shock front and the ejecta, where solar wind pile-up and shock interactions amplify magnetic field fluctuations \citep{2017_Kilpua}. The fluctuation power in both solar wind and ICME regions decreases with distance, reflecting a weakening of magnetic fluctuations in these regions away from the Sun.

The cross-field analysis reveals that at all studied heliocentric distances both sheath and ME regions exhibit significantly reduced $v$--$b$ correlations compared to the surrounding solar wind, which maintains strong Alfv\'{e}nic correlations. The reduced correlations in the sheath may reflect its origin as compressed solar wind plasma retaining residual Alfv\'{e}nic signatures, which are subsequently diminished by locally generated, balanced turbulence driven by velocity shear or shock interactions \citep{2022_Good}. In contrast, the ME regions show very poor correlations (with one exceptional event), consistent with their closed magnetic topology that supports bidirectional propagation of Alfv\'{e}nic fluctuations, and the resulting balance between sunward and anti-sunward wave populations may reduce directional asymmetry and weaken the observed correlation \citep{2020b_Good}.

Spectral fits of magnetic power spectra in the inertial range reveal that turbulence in solar wind regions exhibits variable inertial-range scaling. This includes either a single power law close to $f^{-3/2}$ or $f^{-5/3}$, or a coexistence of both. The non-Alfv\'{e}nic solar wind  intervals (i.e., SSW) are characterized by a single power-law spectrum close to $f^{-5/3}$, with the spectra steepening with increasing heliocentric distance. In contrast, the Alfv\'{e}nic solar wind intervals (ASSW and FSW) display more complex behavior, showing either a single power law near $f^{-3/2}$ or $f^{-5/3}$, or a double power-law form with a transition from $f^{-3/2}$ at lower frequencies to $f^{-5/3}$ at higher frequencies, without any clear dependence on heliocentric distance. These results suggest that Alfv\'{e}nicity may influence the inertial-range scaling in the solar wind. The $f^{-5/3}$ regime likely corresponds to isotropic Kolmogorov turbulence or anisotropic MHD turbulence with weak $\mathbf{v}$--$\mathbf{b}$ alignment in the non-Alfv\'{e}nic regime of solar wind fluctuations, whereas the $f^{-3/2}$ regime is consistent with anisotropic turbulence under strong v--b alignment \citep{Kolmogorov_1941,1995_Goldreich,2006_Boldyrev}. The observed double power-law behavior in Alfv\'{e}nic wind intervals may indicate a transition from a highly imbalanced turbulence state ($|z^{+}|^2 \gg |z^{-}|^2$ or vice versa) at larger scales to a more balanced state ($|z^{+}|^2 \sim |z^{-}|^2$) at smaller scales \citep{2025_Shiladittya}. This finding aligns with recent results by \citet{2025_Shiladittya}, who reported two distinct sub-regimes in Alfv\'{e}nic solar wind and highlighted that Alfv\'{e}nicity might be a key factor in producing such inertial-range breaks, though other effects such as solar wind expansion, geometry, latitude, and turbulence amplitude may also contribute.

Unlike the variable spectral characteristics of the ambient solar wind, both the ICME sheath and ME regions consistently exhibit a single power-law scaling close to $f^{-5/3}$ across all studied distances, independent of speed or Alfv\'{e}nicity. Similar slopes have been reported at 1~au in both sheath and flux rope regions \citep{2021a_Kilpua,2024_Shaikh}. Our results extend these observations into the inner heliosphere, supporting the presence of a well-developed turbulent cascade within ICME substructures. The findings for ICME ME are consistent with \citet{2023_Good}, who also reported a typical spectral index near $-5/3$ in ICME ejecta between 0.25 and 0.95~au. The observed $f^{-5/3}$ scaling may be linked to the reduced $\mathbf{v}$–$\mathbf{b}$ correlations in ICME substructures, reflecting the low cross-helicity values commonly found in ICMEs \citep{2023_Soljento}. However, we report one exceptional event in which the ME exhibits a single power-law spectrum with a slope of $-3/2$ in the inertial range, likely associated with anisotropic turbulence arising from strong $\mathbf{v}$--$\mathbf{b}$ alignment \citep{1995_Goldreich, 2006_Boldyrev}. Notably, this event is consistent with the scenario proposed by \citet{2020b_Good}, in which ICME legs exhibit stronger $\mathbf{v}$--$\mathbf{b}$ coupling and elevated $|\sigma_c|$. The exceptional case therefore provides direct observational evidence that leg encounters can reveal regions of strong Alfvénicity, producing a shallower $f^{-3/2}$ spectrum in place of the canonical $f^{-5/3}$ cascade in the ejecta, demonstrating that the turbulence properties of an ICME ME can depend on the spacecraft's sampling geometry.

We further find that the spectral break separating the inertial and kinetic ranges occurs at higher frequencies (smaller scales) in the sheath and ME compared to the ambient solar wind, suggesting distinct microphysical plasma conditions in these regions. The average break frequencies for the solar wind, sheath, and ME are  $0.53 \pm 0.35$, $1.87 \pm 1.46$, and $1.46 \pm 1.28$~Hz, respectively.  Across all regions, the break frequency decreases with increasing heliocentric distance (Figure~\ref{fig:bp_radial}). The breakpoints tend to lie close to the ion inertial length in both the solar wind and ICME substructures (sheath and ME), with this correspondence being especially pronounced in the low-\( \beta \) ME regions.

Based on these findings, we suggest that magnetic fluctuation power, inertial-range scaling, spectral break location, and the $v$--$b$ correlation can serve as effective indicators for identifying ICME boundaries. In the ambient solar wind, the inertial-range spectrum is highly variable and may be influenced by the level of Alfvénicity, whereas within ICMEs we predominantly observe a single, well-developed turbulent cascade exhibiting a comparatively stable $f^{-5/3}$ slope. Recent studies have demonstrated the feasibility of automatic ICME detection using machine-learning and deep-learning techniques applied to in-situ plasma and magnetic-field measurements (e.g., \citealt{2019_Nguyen, 2022_Rudisser, 2025_Rudisser}). The turbulence-based signatures identified here are complementary to such approaches and could serve as additional, physically interpretable features for training or validating automated ICME identification schemes. 

In real time, only the leading portion of an ICME is observed before the structure fully passes a spacecraft. While a complete spectral characterization of the inertial and kinetic ranges requires sufficiently long time intervals, our results indicate that key transitions, such as enhanced fluctuation power, reduced $v$--$b$ correlation, and an upward shift of the spectral break, already emerge at the sheath onset and during the early ejecta phase. This suggests that spectral diagnostics can be applied progressively using sliding or adaptive time windows, rather than requiring the full ICME interval, making them potentially suitable for near-real-time analyses.
Furthermore, turbulence diagnostics can be used to validate or refine candidate ICME substructure boundaries proposed by machine-learning or deep-learning algorithms by systematically shifting boundary locations and assessing changes in fluctuation power, correlation properties, and break-scale behavior across these intervals. Specifically, the onset of the ICME sheath is marked by an increase in fluctuation power, a shift of the break scale to higher frequencies, and reduced $v$--$b$ correlation. Upon the arrival of the ejecta, the fluctuation power decreases relative to the sheath, the break remains at a higher frequency, and the $v$--$b$ correlation weakens further. Following the ICME passage, the break frequency shifts back toward lower values, and the $v$--$b$ correlation increases, as typically observed in the solar wind. An iterative refinement of boundary placement based on such spectral signatures may further improve boundary identification.

For these diagnostics to be effective, magnetic-field measurements with cadences of order 10~Hz are required to reliably resolve inertial-range fluctuations and identify spectral breaks near ion-kinetic scales. This requirement is met by current missions such as \emph{SolO}, \emph{Parker Solar Probe (PSP)}, and \emph{Wind}. Lower cadences may still capture changes in fluctuation power but would limit the robustness of spectral-break identification, particularly in near-Sun observations. While the present study provides new insights into the turbulent evolution of ICMEs and their distinction from the surrounding solar wind, a larger statistical investigation is required to assess the robustness of these diagnostics across different heliocentric distances, solar wind conditions, and phases of the solar cycle. In future work, this analysis can be extended to a broader set of SolO events, including those observed after March 2024, and to ICME observations from PSP, thereby enabling a more comprehensive investigation of ICME turbulence in the inner heliosphere.

\section*{Acknowledgments}
\begin{acknowledgments}
S. B. acknowledges financial support from STC ISRO project (STC/PHY/2023664O). We acknowledge the use of the ICME catalog hosted at \url{https://helioforecast.space/icmecat}, developed by Christian M\"ostl and collaborators as part of the HELIO4CAST project. We also thank the Solar Orbiter mission team and data providers for making the in-situ measurements used in this study publicly available. We express our gratitude to the anonymous referee for providing us with valuable comments.
\end{acknowledgments}

\section*{Data Availability}
All data used in our study is publicly accessible via NASA CDAWeb (\href{https://cdaweb.gsfc.nasa.gov}{https://cdaweb.gsfc.nasa.gov}).

\begin{acknowledgments}

\end{acknowledgments}

\appendix
\renewcommand{\thefigure}{A\arabic{figure}}
\setcounter{figure}{0}  
\renewcommand{\thetable}{A\arabic{table}}
\setcounter{table}{0}

\section{Tables}\label{tab:appendix_table}


\begin{table*}[ht]
\centering
\renewcommand{\arraystretch}{1.2}
\caption{List of ICME Events Analyzed}
\label{tab:icme_events}
\begin{tabular}{c c c c c c c c c c}
\hline
Event & \multicolumn{1}{c}{ICME} & \multicolumn{1}{c}{ME} & \multicolumn{1}{c}{ICME/ME} &  \multicolumn{1}{c}{R} & \multicolumn{4}{c}{$V_{\mathrm{avg}}$ } \\
Number & start time  & start time & end time & [AU] &\multicolumn{4}{c}{$[\mathrm{km\,s^{-1}}]$ } \\ 
 & [UT] & [UT] & [UT] &  & SW1 & Sheath & ME & SW2 \\
\hline
\hline
1  & 2023-04-10 04:34 & 2023-04-10 10:12 & 2023-04-10 13:56& 0.29 & 336 & 532 & 519 & 383 \\
2  & 2023-10-10 22:32 & 2023-10-11 03:27 & 2023-10-11 09:19& 0.30 & 372 & 764 & 572 & 371 \\
3  & 2023-10-17 10:24 & 2023-10-17 17:46 & 2023-10-18 11:50& 0.37 & 332 & 558 & 433 & 358 \\
4  & 2023-03-21 12:23 & 2023-03-21 13:24 & 2023-03-21 19:49& 0.50 & 651 & 638 & 558 & 516 \\
5  & 2023-11-11 09:14 & 2023-11-12 03:57 & 2023-11-12 20:21& 0.69 & 362 & 467 & 378 & 370 \\
6  & 2021-10-14 23:12 & 2021-10-15 09:54 & 2021-10-15 23:11& 0.72 & 307 & 312 & 332 & 329 \\
7  & 2023-02-17 14:03 & 2023-02-17 23:23 & 2023-02-18 05:18& 0.82 & 365 & 454 & 458 & 393 \\
8  & 2021-11-03 14:03 & 2021-11-04 01:25 & 2021-11-04 19:47& 0.85 & 459 & 675 & 645 & 540 \\
9  & 2021-05-06 18:26 & 2021-05-06 18:26 & 2021-05-07 15:07& 0.91 & 408 & \ldots & 384 & 363 \\
10 & 2021-05-10 03:55 & 2021-05-10 10:01 & 2021-05-11 14:55& 0.92 & 378 & 548 & 502 & 422 \\
11 & 2022-07-25 06:24 & 2022-07-25 11:45 & 2022-07-26 02:29& 0.98 & 341 & 872 & 814 & 502 \\
12 & 2022-06-28 08:10 & 2022-06-28 12:12 & 2022-06-28 19:39& 1.01 & 433 & 635 & 644 & 562 \\
\hline
\end{tabular}

\end{table*}

\begin{figure*}
    \centering
    \includegraphics[width=0.3\textwidth]{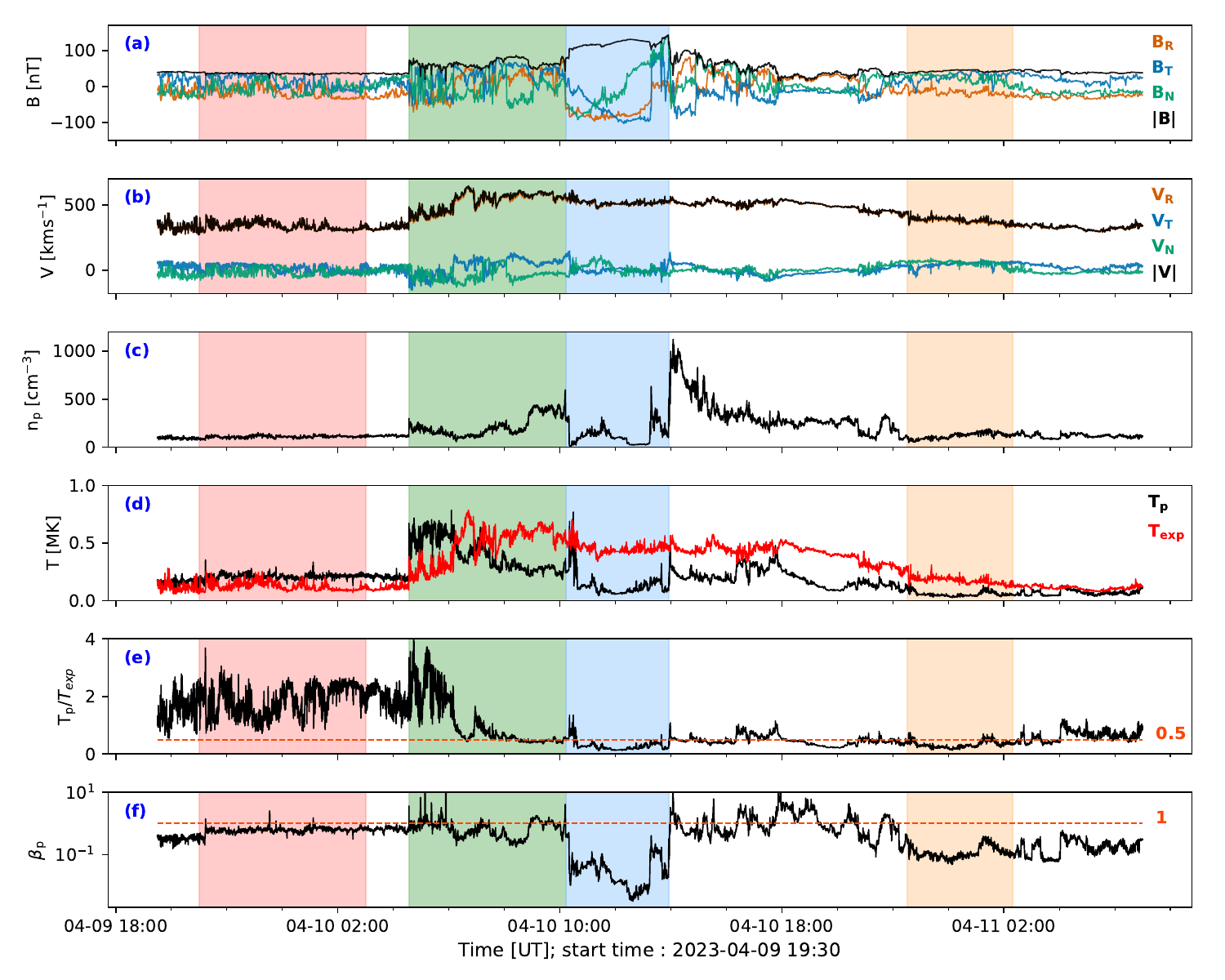} \hfill
    \includegraphics[width=0.3\textwidth]{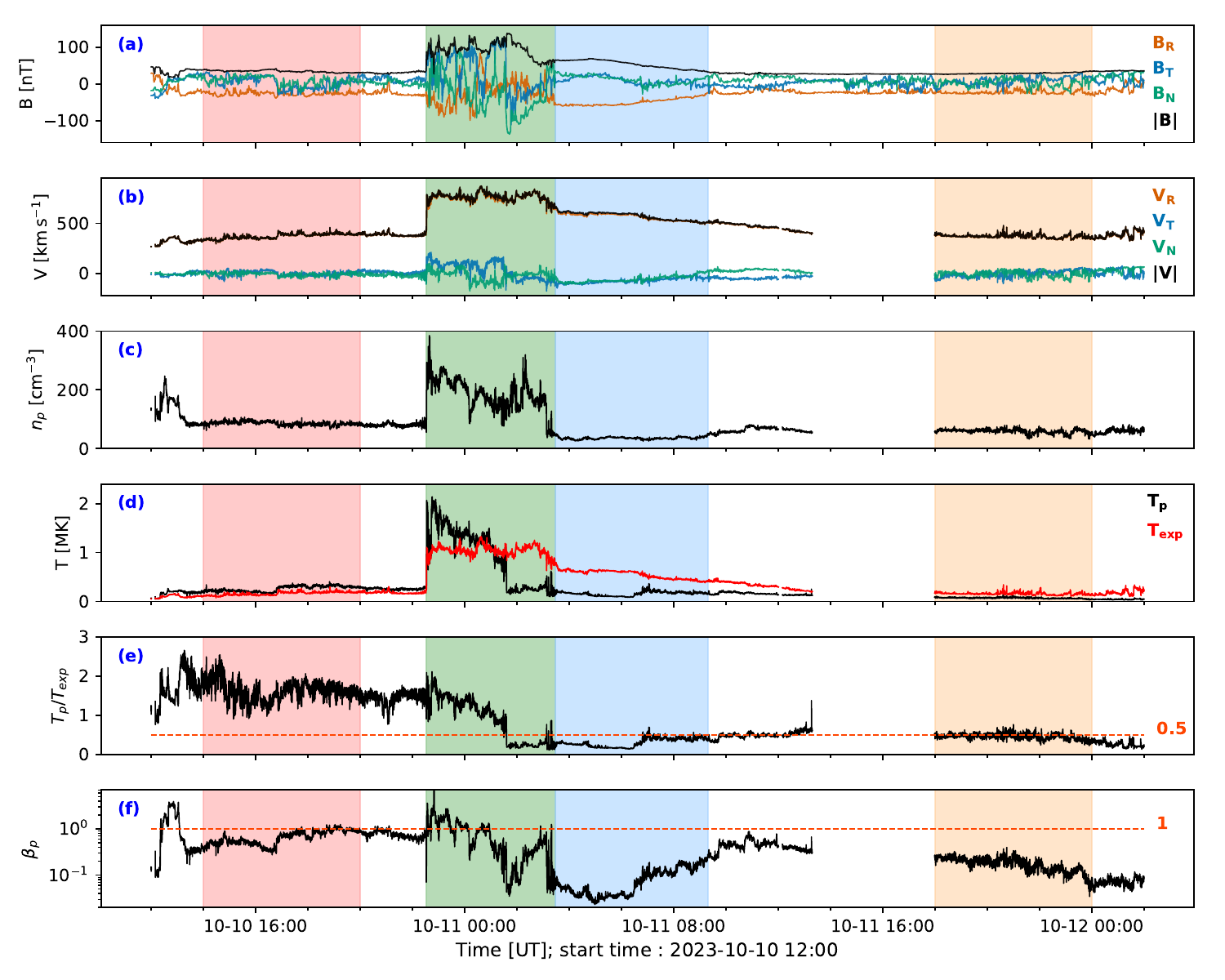} \hfill
    \includegraphics[width=0.3\textwidth]{figures/SolO_icme_20231017.pdf} \\
    \vspace{1em}
    \includegraphics[width=0.3\textwidth]{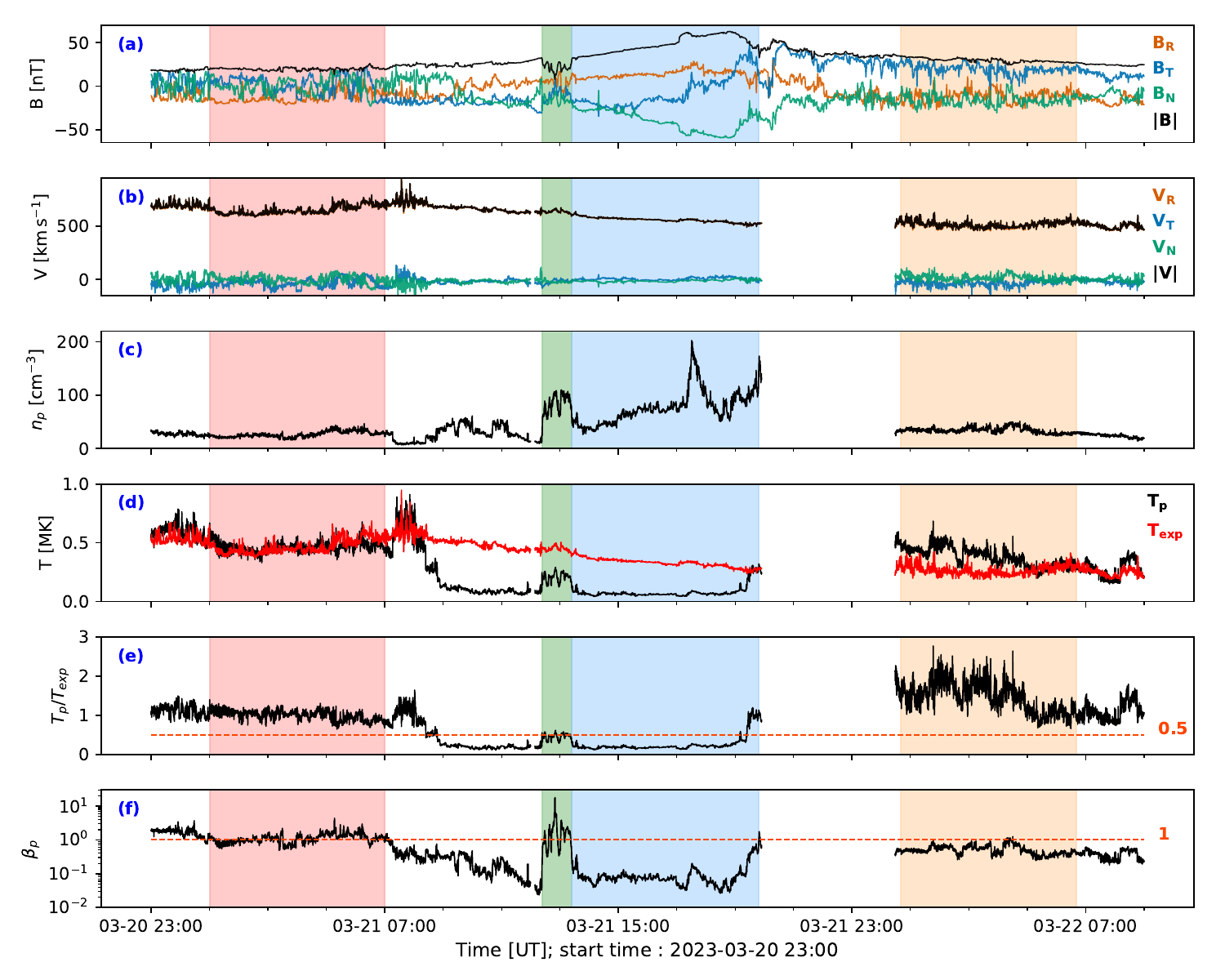} \hfill
    \includegraphics[width=0.3\textwidth]{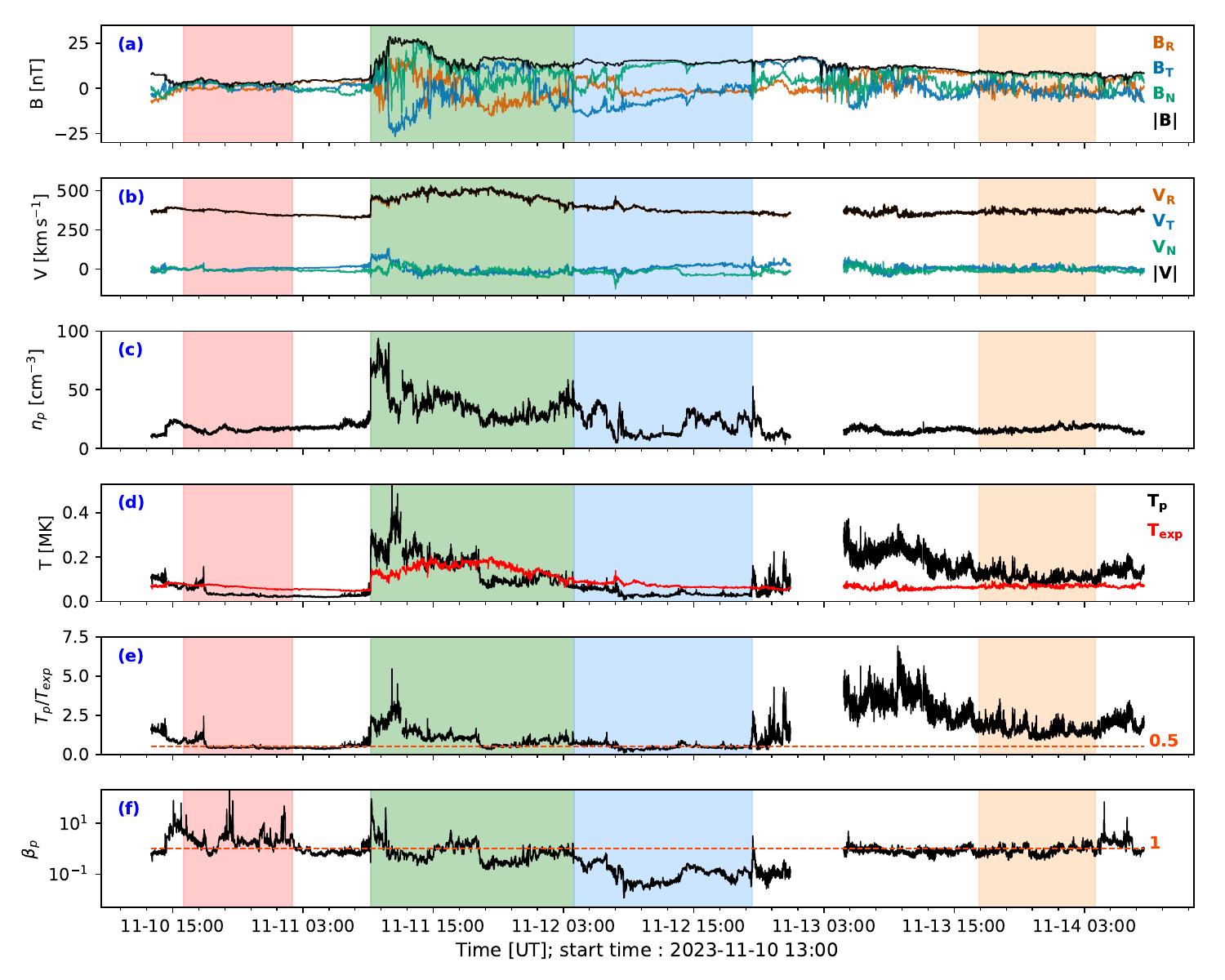} \hfill
    \includegraphics[width=0.3\textwidth]{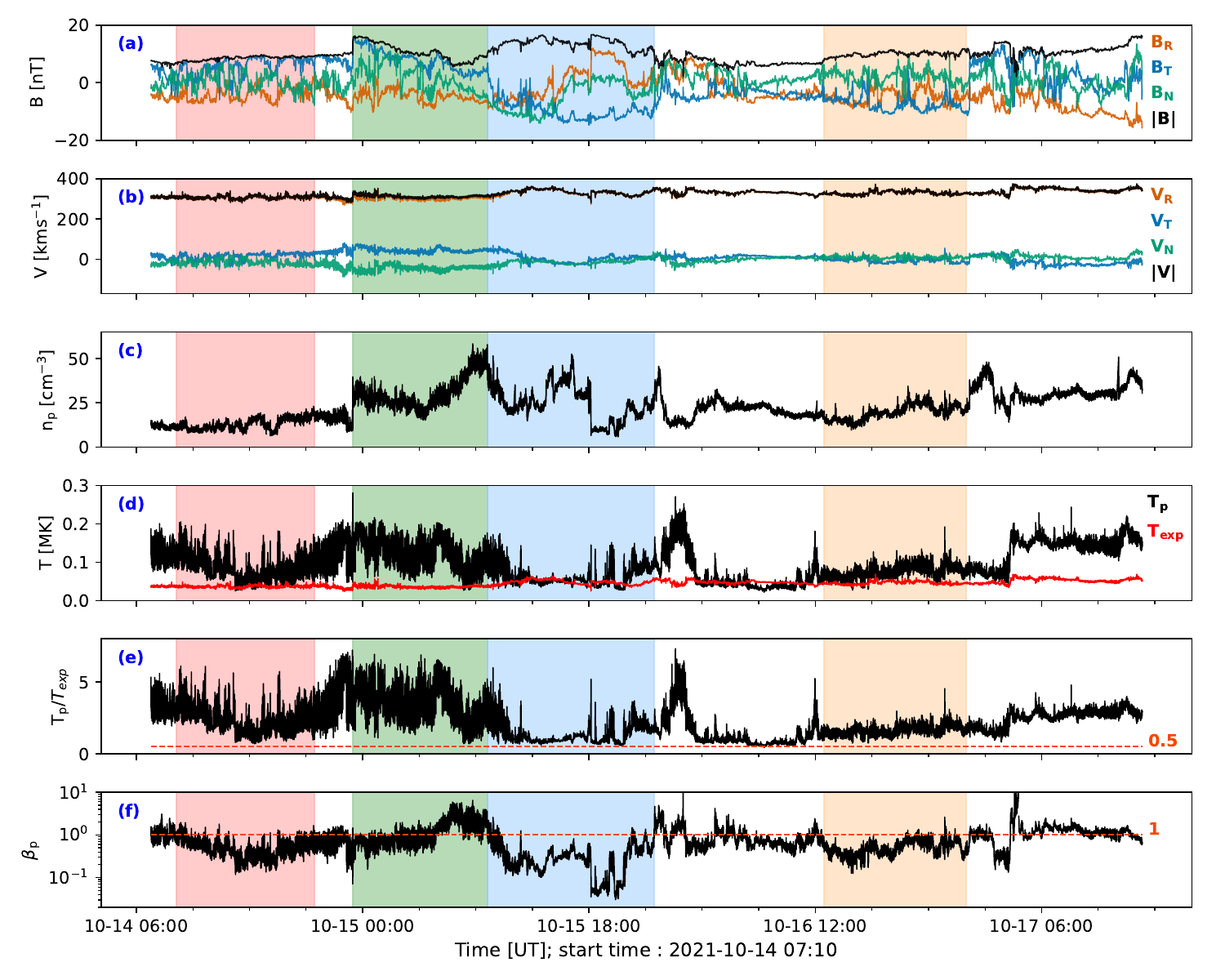} \\
    \vspace{1em}
    \includegraphics[width=0.3\textwidth]{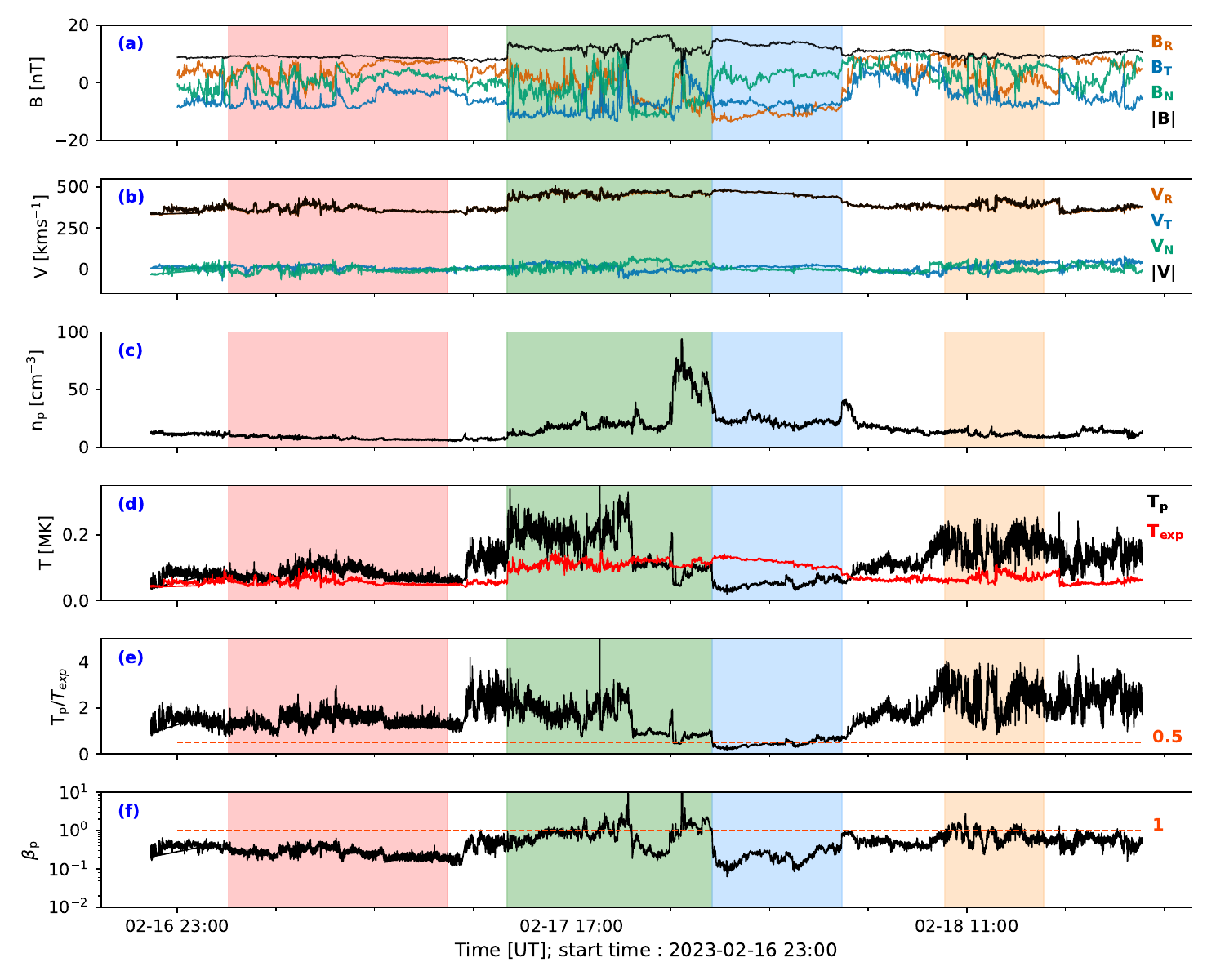} \hfill
    \includegraphics[width=0.3\textwidth]{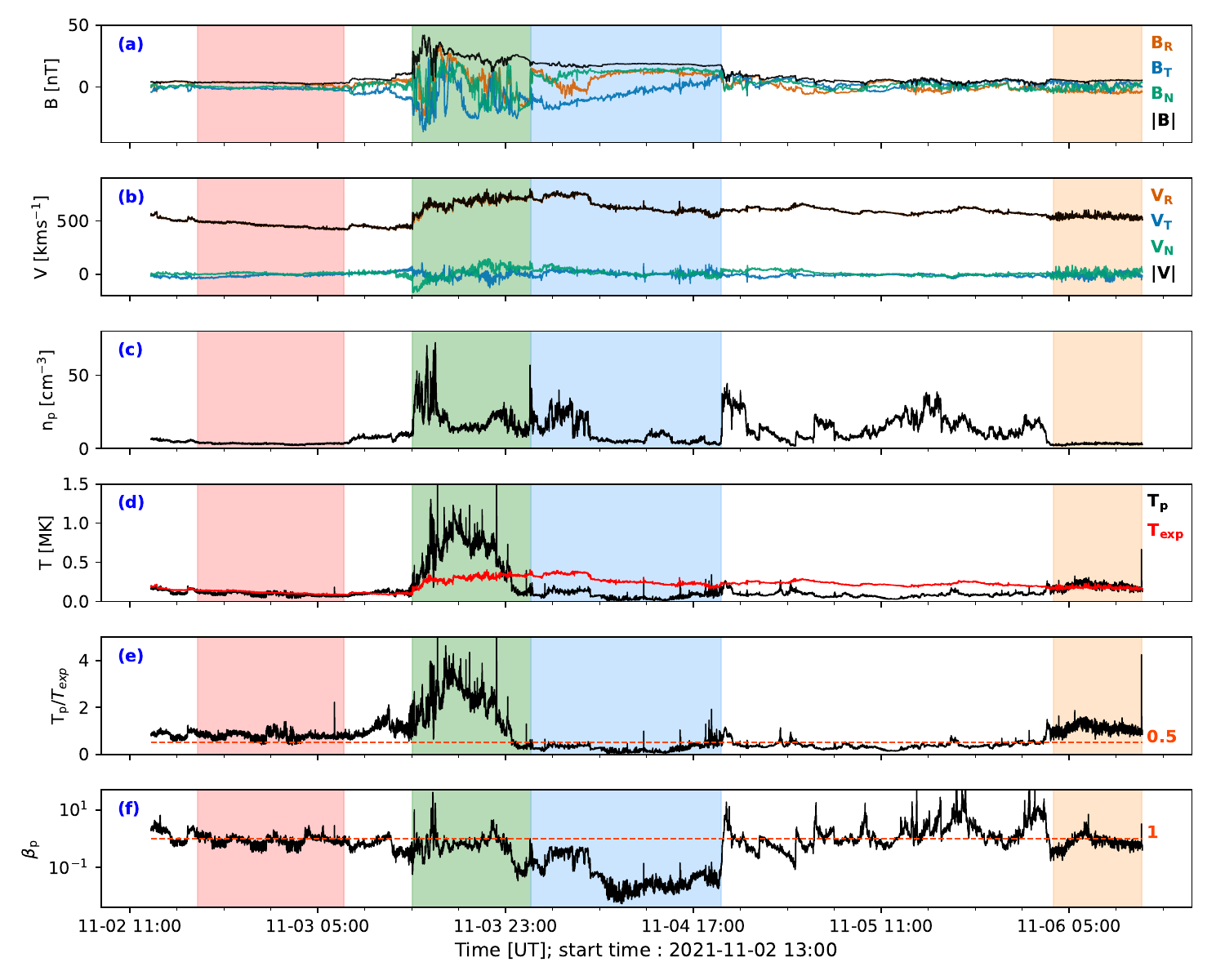} \hfill
    \includegraphics[width=0.3\textwidth]{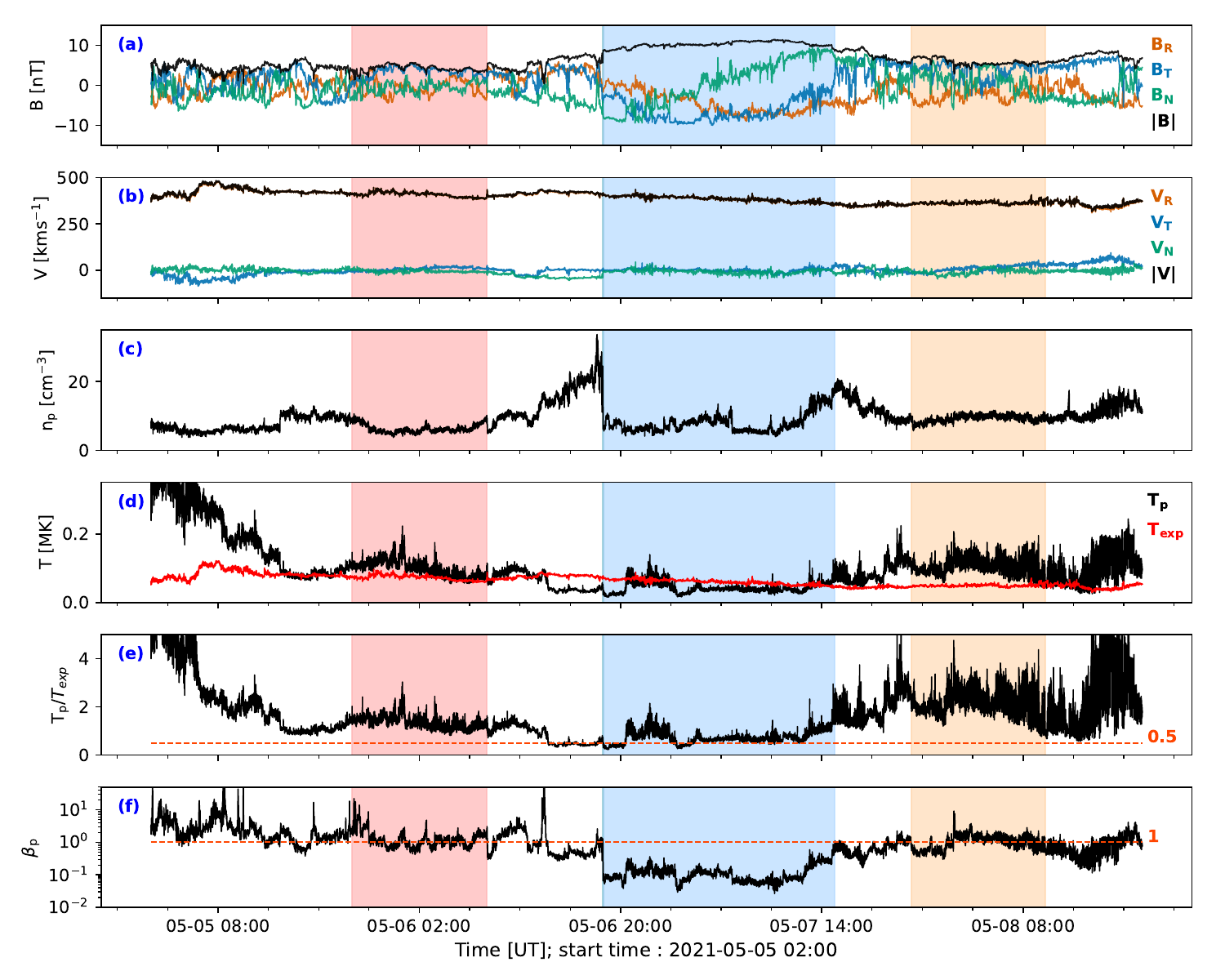}  \\
     \vspace{1em}
    \includegraphics[width=0.3\textwidth]{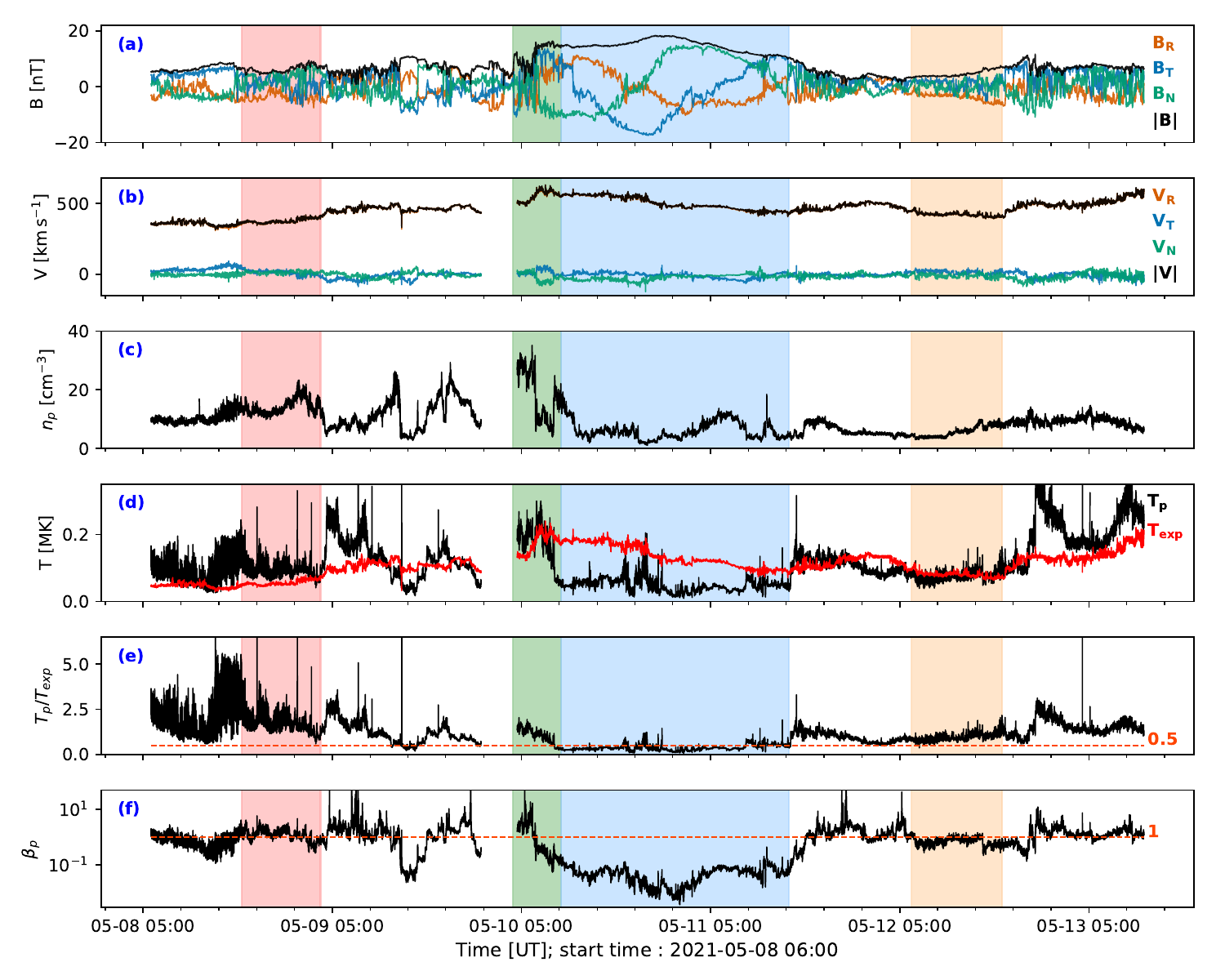} \hfill
    \includegraphics[width=0.3\textwidth]{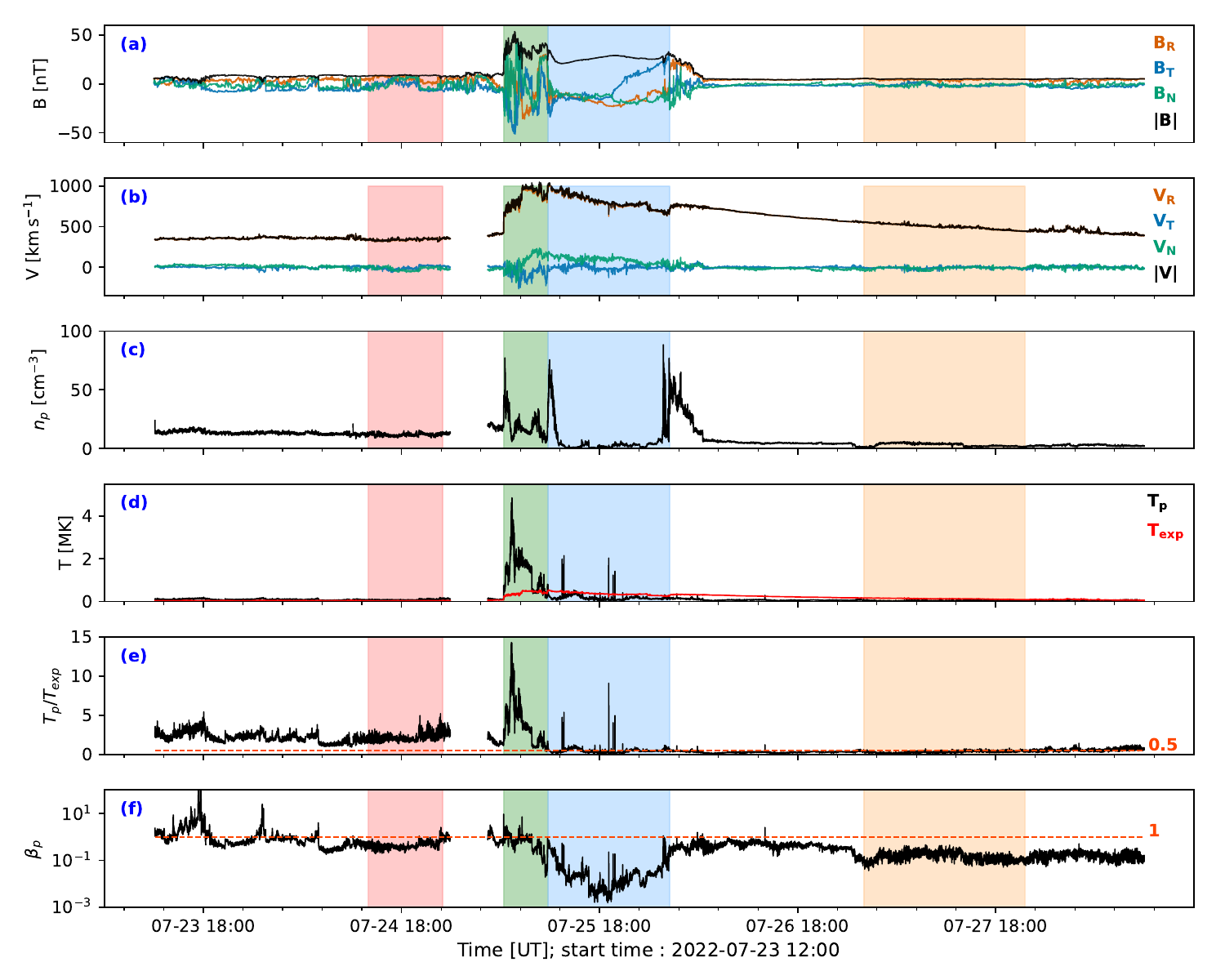} \hfill
    \includegraphics[width=0.3\textwidth]{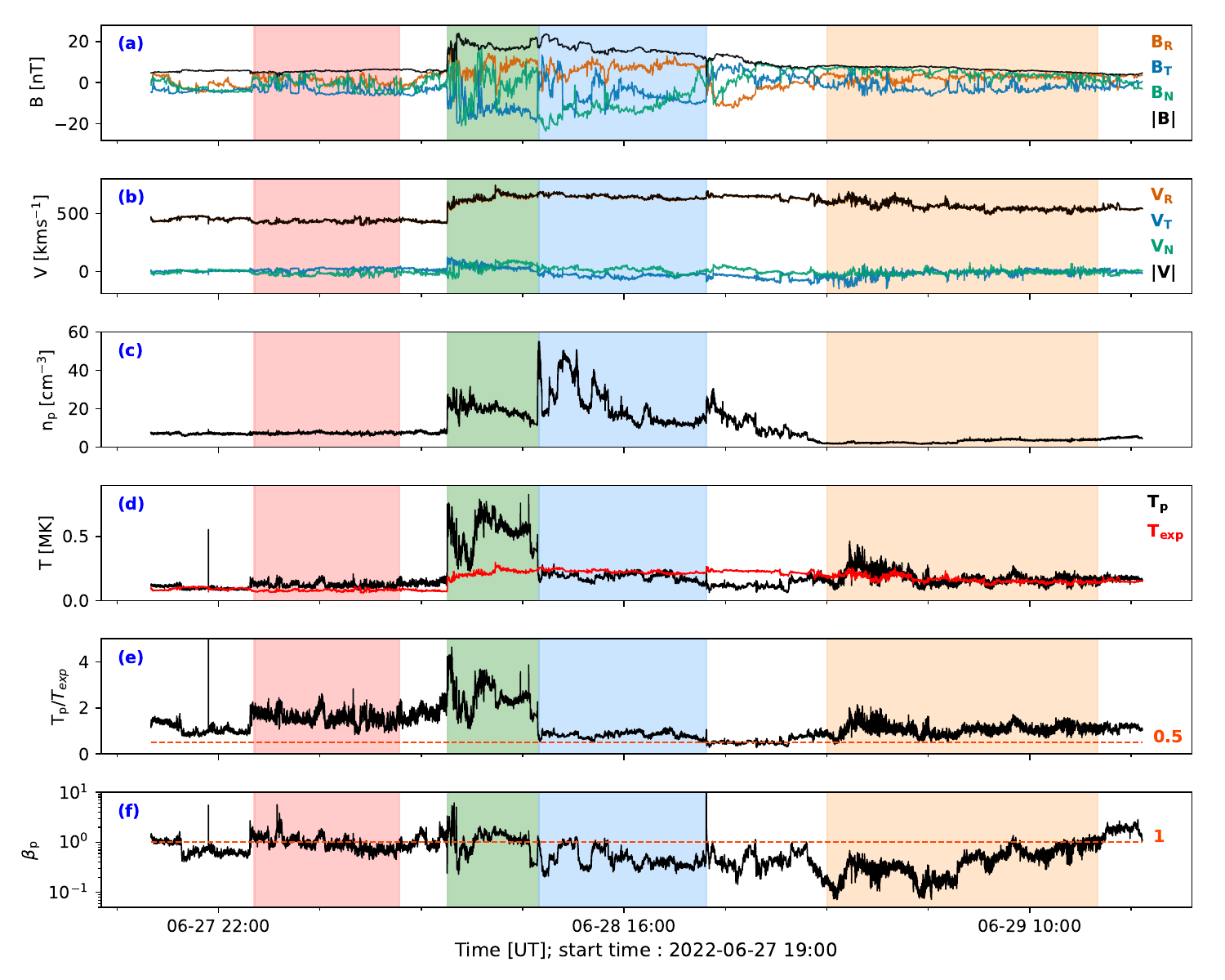} \\

    \caption{Magnetic field and plasma data for all ICMEs analyzed in this study. Panel details are the same as in Figure~\ref{fig:icme_231017}. Color-shaded regions indicate SW1 (light salmon pink), sheath (light green), ME (light blue), and SW2 (light peach).}
\label{fig:rawdata_a}
\end{figure*}

\begin{table*}[ht]
\centering
\renewcommand{\arraystretch}{1.2}

\caption{
\begin{minipage}[t]{0.8\linewidth}
Classification of solar wind intervals based on speed and the correlation coefficient (\texttt{cc\_r}) between magnetic and velocity field fluctuations: \\[2pt]
(A) SSW: speed $\leq$ 500 km~s$^{-1}$ and \texttt{cc\_r} $<$ 0.6 \\
(B) FSW: speed $>$ 500 km~s$^{-1}$ and \texttt{cc\_r} $\geq$ 0.6 \\
(C) ASSW: speed $\leq$ 500 km~s$^{-1}$ and\texttt{cc\_r} $\geq$ 0.6
\end{minipage}
}

\label{tab:sw_class}
\begin{tabular}{c c c c c}
\hline
\textbf{Type} & \textbf{Event} & \textbf{Interval} & \textbf{Speed}
 & \textbf{Inertial Range Slope} \\
&  &  & {\boldmath$[\mathrm{km\,s^{-1}}]$} &  \\
\hline
\hline
SSW & 5 & SW1 & 362 & –1.66 \\
    & 8 & SW1 & 459 & – \\
    & 9 & SW1 & 408 & –1.77 \\
\hline
FSW & 4  & SW1 and SW2 & 651 and 516 & –1.65 and –1.51 \\
    & 8  & SW2          & 540         & –1.69 \\
    & 11 & SW2          & 502         & –1.70 \\
    & 12 & SW2          & 562         & (–1.48, –1.74) \\
\hline
ASSW & 1  & SW1 and SW2 & 336 and 383 & (–1.58, –1.73) and –1.51 \\
     & 2  & SW1 and SW2 & 372 and 371 & (–1.51, –1.65) and –1.62 \\
     & 3  & SW1 and SW2 & 332 and 358 & (–1.57, –1.82) and –1.55 \\
     & 5  & SW1 and SW2 & 362 and 370 & –1.66 and –1.60 \\
     & 6  & SW1 and SW2 & 307 and 329 & –1.54 and –1.62 \\
     & 7  & SW1 and SW2 & 365 and 393 & –1.49 and –1.61 \\
     & 9  & SW2         & 363         & –1.65 \\
     & 10 & SW1 and SW2 & 378 and 422 & (–1.59, –1.71) and –1.67 \\
     & 11 & SW1         & 341         & –1.67 \\
     & 12 & SW1         & 433         & –1.65 \\
\hline
\end{tabular}
\end{table*}

\begin{figure*} 
    \centering
    \includegraphics[width=\textwidth]{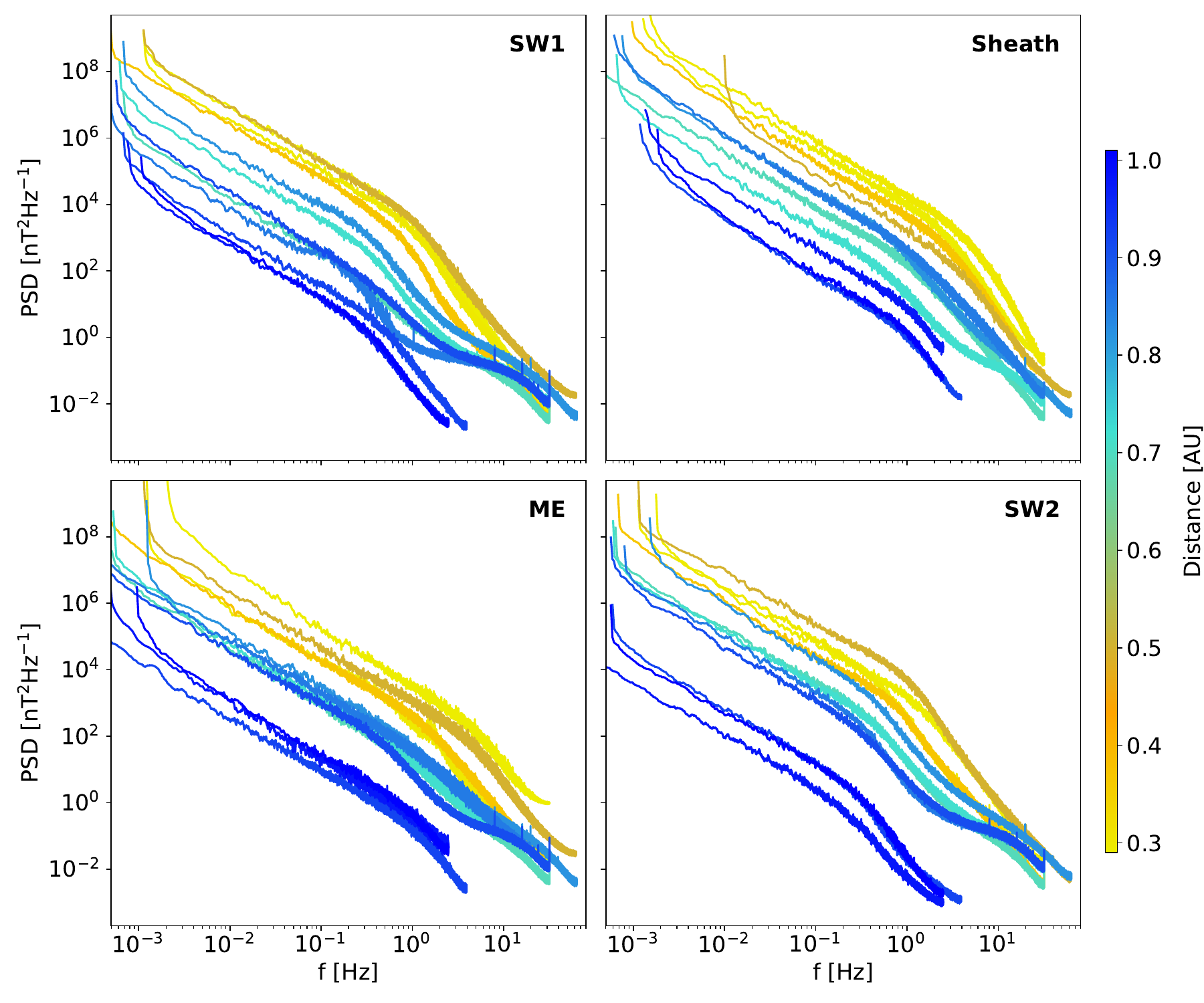}
    \caption{Smoothed trace PSDs of magnetic field fluctuations for the selected regions (SW1, sheath, ME, and SW2) across all analyzed ICMEs. Spectra are color-coded by radial distance.}
    \label{fig:psds_all}
\end{figure*}

\begin{figure*} 
    \centering
    \includegraphics[width=0.9\textwidth]{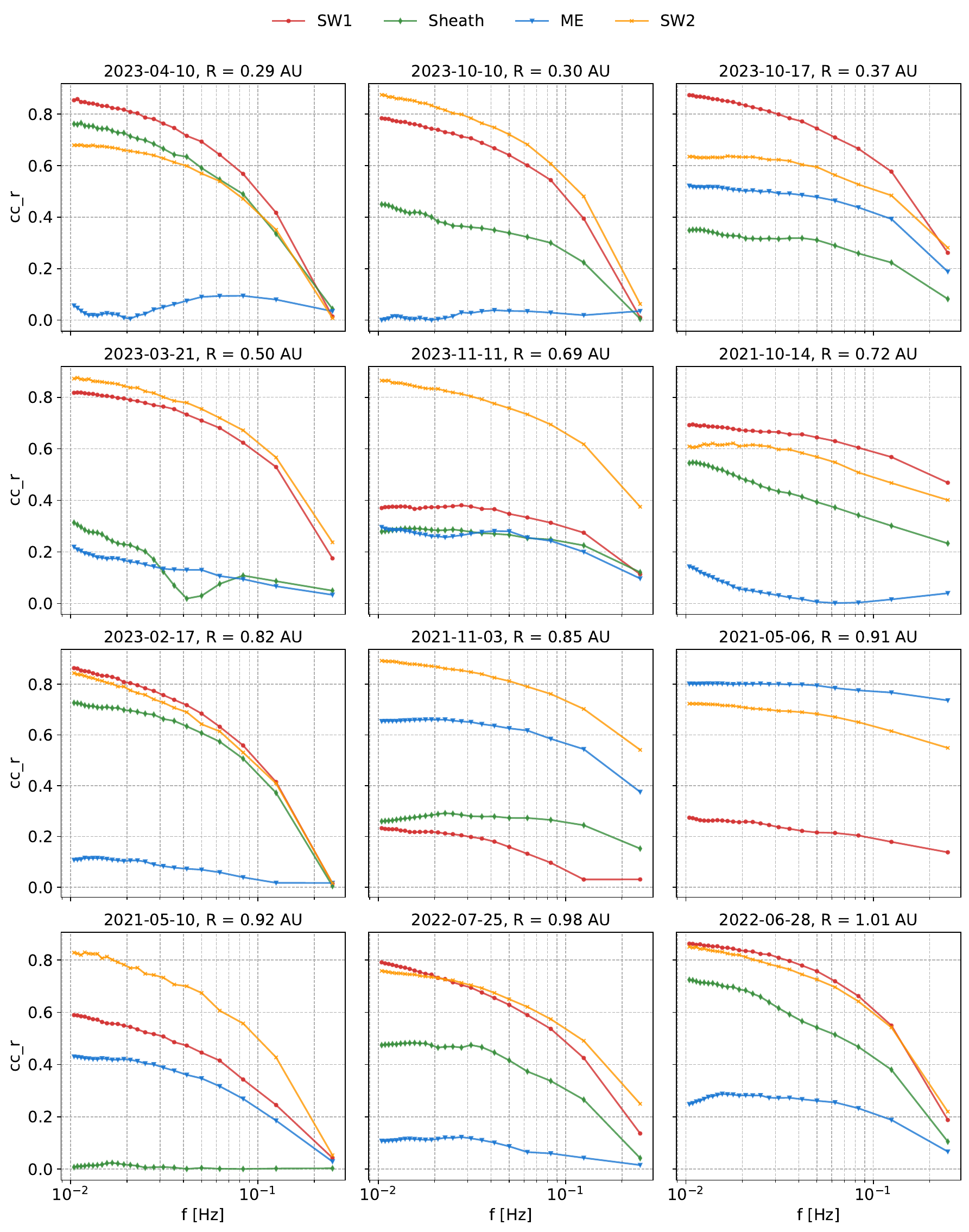}
    \caption{Absolute cross-correlation coefficients between magnetic and velocity field fluctuations in the R component, plotted as a function of frequency ($1/\tau$), for the SW1, sheath, ME, and SW2 regions across all ICME events. Plots are color-coded by region.}
    \label{fig:cc_all}
\end{figure*}



\end{document}